\documentclass[twocolumn]{aastex631}

\let\tablenum\relax

\usepackage{graphicx}
\usepackage{booktabs,array}
\usepackage{siunitx}
\usepackage{amsmath,amssymb}
\usepackage{microtype}
\usepackage[T1]{fontenc}
\usepackage[utf8]{inputenc}
\usepackage{lmodern}
\usepackage{placeins}

\renewcommand{\arraystretch}{0.70}



% ...

\newcommand{\um}{\micron\;}

\newcommand{\Msun}{\hbox{M$_\odot$}}

\newcommand{\z}{$1<z\lesssim2$\;}

\newcommand{\no}{\nodata}
\newcommand{\Msol}{\ensuremath{\mathrm{M}_\sun}}

\newcommand*\cleartoleftpage{%
  \clearpage
  \ifodd\value{page}\hbox{}\newpage\fi
}

\def \m {MAGPHYS\;}

\def\nar{\ref@jnl{New A Rev.}}          

\shorttitle{Bolometric Luminosity Correction Recipe 
}
\shortauthors{Azadi et al.}

\begin{document}

\title{A Bolometric Luminosity Correction Recipe for AGN at Any Epoch}

\author[0000-0001-6004-9728]{Mojegan Azadi}
\affiliation{Center for Astrophysics $|$ Harvard \& Smithsonian, 60 Garden Street, Cambridge, MA, 02138, USA}
\author[0000-0003-1809-2364]{Belinda Wilkes}
\affiliation{Center for Astrophysics $|$ Harvard \& Smithsonian, 60 Garden Street, Cambridge, MA, 02138, USA}
\author[0000-0001-5513-029X]{Joanna Kuraszkiewicz}
\affiliation{Center for Astrophysics $|$ Harvard \& Smithsonian, 60 Garden Street, Cambridge, MA, 02138, USA}
\author[0000-0002-9895-5758]{S. P. Willner}
\affiliation{Center for Astrophysics $|$ Harvard \& Smithsonian, 60 Garden Street, Cambridge, MA, 02138, USA}
\author[0000-0002-3993-0745]{Matthew L.\ N.\ Ashby}
\affiliation{Center for Astrophysics $|$ Harvard \& Smithsonian, 60 Garden Street, Cambridge, MA, 02138, USA}


\begin{abstract}

Understanding how active galactic nuclei (AGN) affect their host galaxies requires determining their total radiative power across all wavelengths (i.e., bolometric luminosities). We show how AGN accretion disk spectral energy distribution (SED) templates, parameterized by supermassive black hole (SMBH) mass, Eddington ratio, spin, and inclination, can be used to estimate total radiated luminosities. Bolometric luminosities are calculated by integrating the accretion disk SEDs from 1$\mu$m to 10keV over $0^\circ$--$90^\circ$ inclinations, ensuring consistent treatment of wavelength gaps, avoiding double-counting reprocessed emission, and accounting for anisotropy of visible--UV emission at different inclinations. The SED, and resulting bolometric corrections, depend strongly on SMBH mass and Eddington ratio, but only weakly on spin and inclination. Increasing SMBH mass produces cooler disks peaking at lower frequencies, whereas higher Eddington ratios (and spins) yield hotter disks peaking at higher frequencies. Larger inclinations suppress the visible--UV portion of the SED, whereas X-ray emission remains nearly isotropic. Bolometric corrections in the visible--NUV range (5100\AA-3000\AA) show strong dependence on SMBH mass, while X-ray bolometric corrections depend strongly on the Eddington ratio. Near the SED peak (FUV; $\sim$1450\AA), parameter dependencies are weak, making this band particularly robust for estimating bolometric corrections. The X-ray band is reliable, though dependence on the Eddington ratio introduces a wide dynamic range. Because our SEDs are intrinsic and defined in the rest-frame, their application to Type 1 AGN is straightforward. For other AGN, however, corrections for obscuration by the host galaxy and torus are required in many cases.

\end{abstract}


\section{Introduction} \label{sec:intro}


Active Galactic Nuclei (AGN), powered by the accretion of gas and dust onto supermassive black holes (SMBHs) at their centers, are among the most powerful objects in the Universe.  AGN present unique observational signatures 
spanning more than ten decades in wavelength from radio to $\gamma$-ray.     
In practice, observations of individual AGN tend to be restricted to relatively narrow wavelength ranges, affording only limited views of their total radiated power. However, to understand AGN physics and their impact on their environment (i.e., host galaxy, galaxy cluster), it is crucial to determine their total radiative power (i.e., bolometric luminosities) across all wavelengths.

In principle, AGN bolometric luminosities can be estimated by integrating their spectral energy distributions (SEDs) over all electromagnetic frequencies. However, in practice, this is not feasible because observations generally cover limited and discontinuous frequency ranges. To compensate, studies rely on gap repair \citep[e.g.,][]{Elvis1994, Richards2006}. 
An additional complication is that not all AGN emissions are isotropic. While low-frequency radio emission from the lobes is isotropic, mid-infrared (MIR) emission from the torus is less so, and visible--UV emission from accretion disks is strongly anisotropic.
To account for the anisotropic emission, some studies \citep[e.g.,][]{Hubeny2001,R12} consider an average viewing angle and adjust the bolometric luminosity accordingly. 
However, not only is this statistical correction inaccurate for individual objects, but also the simplification is only acceptable in the Newtonian regime.  Relativistic effects, such as aberration and beaming, introduce a stronger viewing-angle dependence in the observed SED \citep[e.g.,][]{Hubeny2001, Nemmen2010, R12}.

\begin{figure*} [h!]
    \centering
    \includegraphics[height=0.99\textwidth,angle =0]{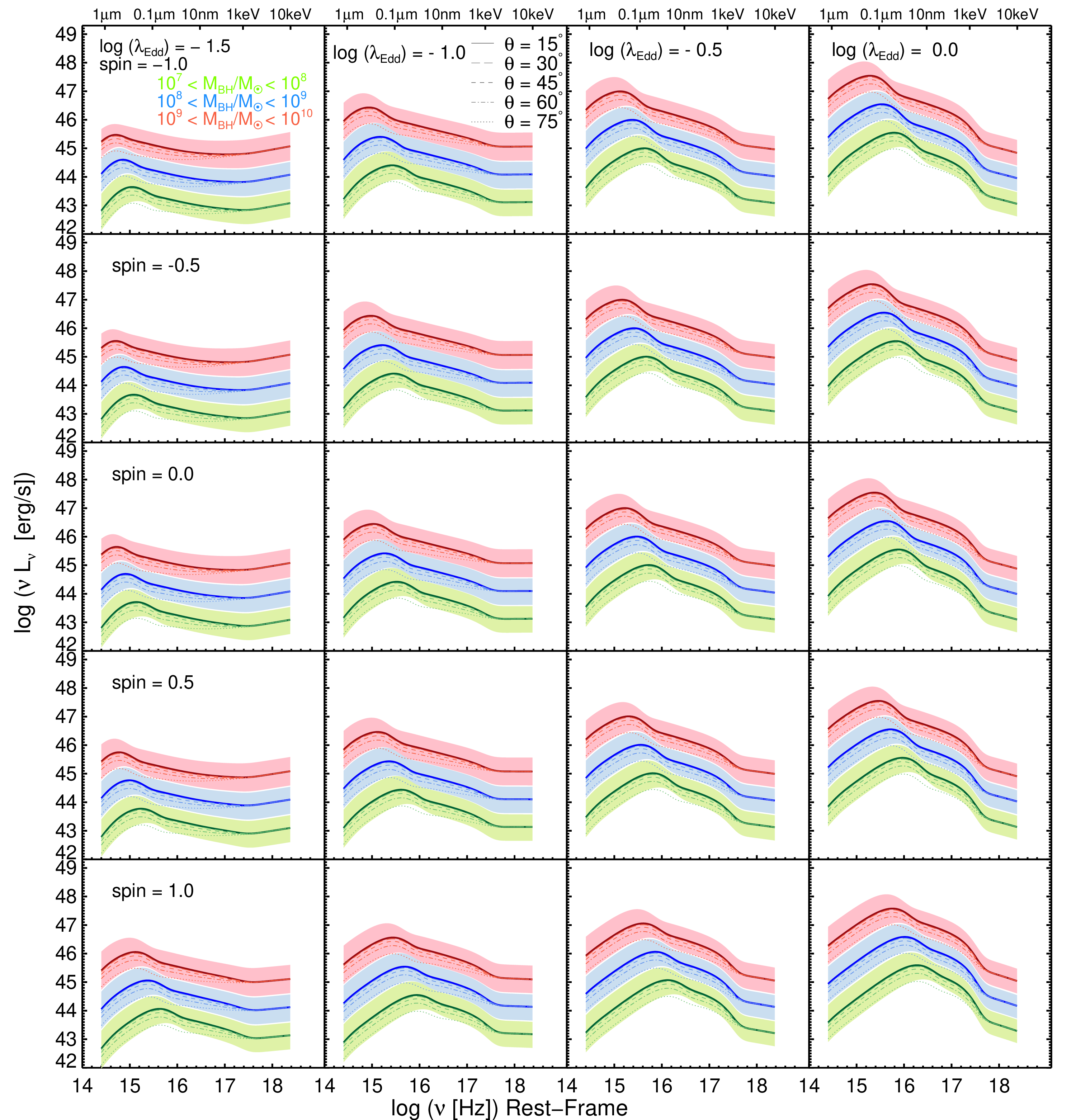}
    \caption{Accretion disk templates created from the \cite{Kubota2018} QSOSED model. In each panel, the SMBH mass varies from $10^{7}$ to $10^{10}$~\Msun\ with each increasing decade represented by green, blue, and pink shaded areas, respectively. The Eddington ratio increases from left to right but is fixed in each panel as labeled. Spin varies from retrograde to prograde from top to bottom. The inclination angle increases from $15^{\circ}$ to $75^{\circ}$ in each panel. The solid lines show the median SED in each mass bin at an inclination angle of $15^{\circ}$, and the shaded regions show the entire range.  
  An increase in SMBH mass results in a cooler disk peaking at lower frequencies, while an increase in Eddington ratio (and spin) results in a hotter accretion disk peaking at higher frequencies. Numerical values are given in Table \ref{tab:app1} in the appendix.}
    \label{fig:new_ad_mass}
\end{figure*}
\begin{figure*} 
    \centering
    \includegraphics[height=0.99\textwidth,angle =0]{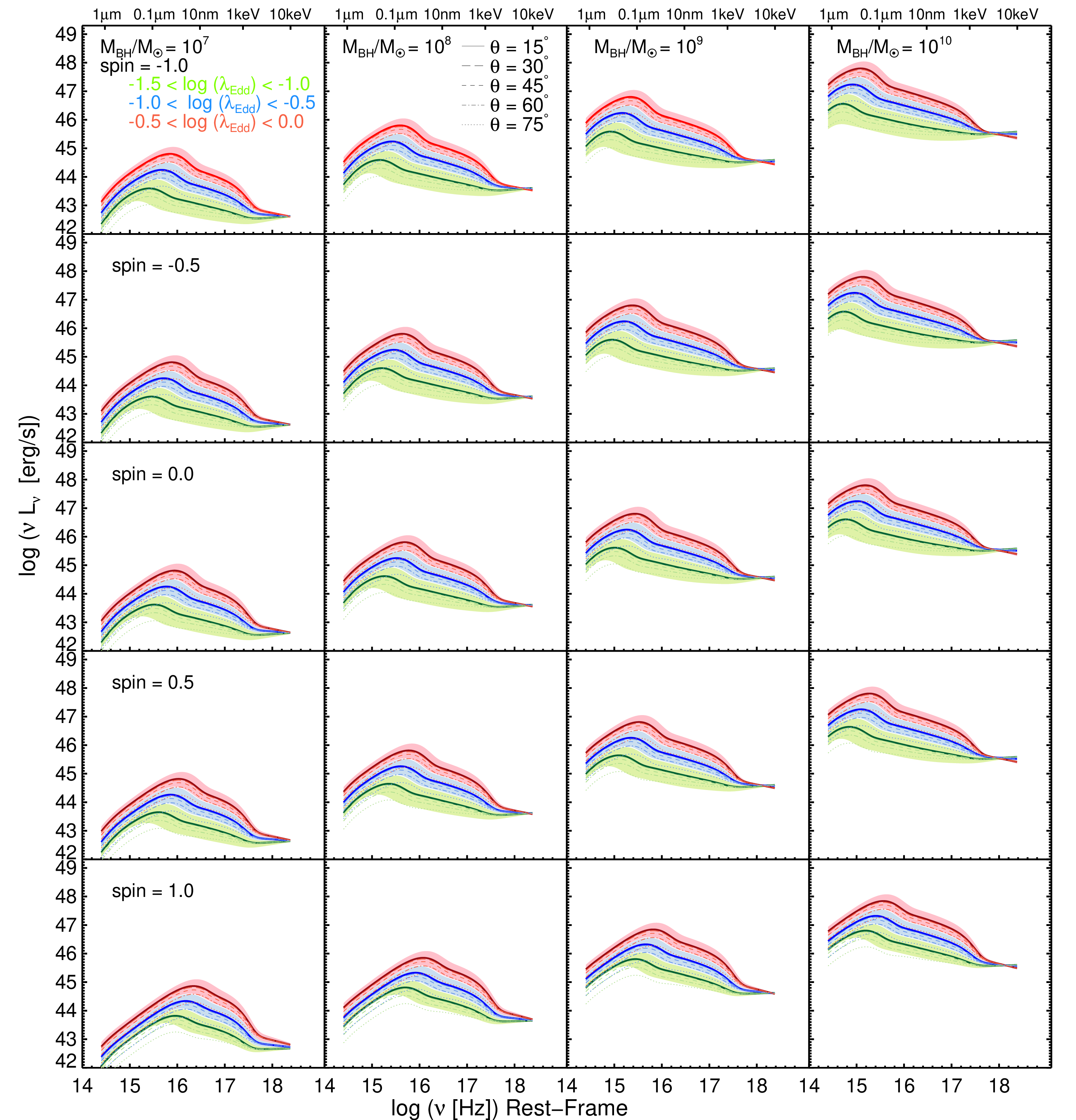}
    \caption{ Accretion disk templates created from the \cite{Kubota2018} QSOSED model. In each panel, the Eddington ratio varies from $\log \lambda_{\mathrm{Edd}} = -1.5$ to $\log \lambda_{\mathrm{Edd}} = 0.0$ with each increasing decade represented by green, blue, and pink shaded areas, respectively. The SMBH mass increases from left to right but is fixed in each panel as labeled. Spin varies from retrograde to prograde from top to bottom. The inclination angle increases from $15^{\circ}$ to $75^{\circ}$ in each panel. The solid lines show the median SED in each Eddington ratio bin at an inclination angle of $15^{\circ}$, and the shaded regions show the entire range. An increase in the Eddington ratio (and spin) results in a hotter disk peaking at higher frequencies, while an increase in SMBH mass results in a cooler accretion disk peaking at lower frequencies. Numerical values are given in Table \ref{tab:app2} in the appendix.}
    \label{fig:new_ad_mdot}
\end{figure*}

The dependence of anisotropic AGN emission on wavelength has led to disagreement as to which wavelength range should be employed 
when estimating an AGN's bolometric luminosity.
Some studies include both the visible--UV emission from the accretion disk and the MIR emission from the torus \citep[e.g.,][]{Elvis1994,Richards2006}. However, others argue that including MIR emission overestimates the bolometric luminosity \citep[e.g.,][]{Marconi2004,Nemmen2010,R12} by double-counting the visible--UV photons which are scattered and re-emitted to power the MIR emission.

Earlier studies determined the bolometric correction from an average SED \citep[e.g.,][]{Elvis1994}. However, no average can be representative of the general AGN population,  so studies moved towards
generating a range of SEDs around the average based on the correlation between the visible-to-X-ray spectral index $\alpha_{\rm ox}$ and 2500\,\AA\ luminosity \citep{V2003} to construct accretion-disk templates \citep[e.g.,][]{Marconi2004,Richards2006,Hopkins2007}. 
Some of these studies found the bolometric correction is luminosity-dependent \citep[e.g.,][]{Marconi2004,Hopkins2007}, while others 
found that the Eddington ratio is the primary determinant of the bolometric correction \citep[e.g.,][]{V2007}.

This paper estimates bolometric corrections for AGN  by integrating model SEDs of accretion disks that represent a general population of AGN\null. We employ the \cite{Kubota2018} QSOSED model, which characterizes the accretion disk emission using four parameters: black hole mass, Eddington ratio, spin, and inclination angle. Integrating the SED templates from  $1\,\mu$m to 10\,keV, varying the parameters over their full expected range, gives the bolometric luminosity and corresponding bolometric correction factors. Additionally, we explore the relationship between the parameters and the shape and normalization of the accretion disk SED and, consequently, the bolometric correction, as a function of wavelength.
The paper is organized as follows: Section~\ref{sec:ad}  describes the \citet{Kubota2018} accretion-disk model and the templates we created. Section~\ref{sec:bc_all} applies the templates to derive bolometric corrections for a general AGN population.   Section \ref{sec:discussion} discusses the findings and compares them with results in the literature. A recipe for observers is presented in Section \ref{sec:recipe} and a summary of our findings is presented in Section~\ref{sec:summary}.

\section{Accretion Disk Model}\label{sec:ad}

\begin{table*}[tb] 
\centering
    \caption {Parameters in the QSOSED Accretion-disk Model 
       }
    \begin{tabular}{ ll l l}
    \hline \hline \vspace{1pt}
         Parameter & Abbreviation & Acceptable range  & Range considered\\
  &  & in QSOSED & in this study\\
         \hline  
SMBH mass & $M_{\rm SMBH}$  &$[10^5,10^{10}]$~\Msol\ &$[10^7,10^{10}] $~\Msol\\
Eddington ratio &$\rm \lambda_{Edd}$  & [$10^{-1.65},10^{0.39}$] & [$10^{-1.5},10^{0.0}$] \\
SMBH spin &$a$ & [$-0.998$,0.998] &[$-0.998$,0.998]  \\
inclination angle & $\theta$ &[$0^{\circ}$,$87^{\circ}$]&[$15^{\circ}$,$75^{\circ}$]\\
\hline
    \end{tabular}
    \label{tab:qsosed}
   \end{table*}

\begin{figure*}[h!]
    \centering
    \includegraphics[height=0.50\textwidth]{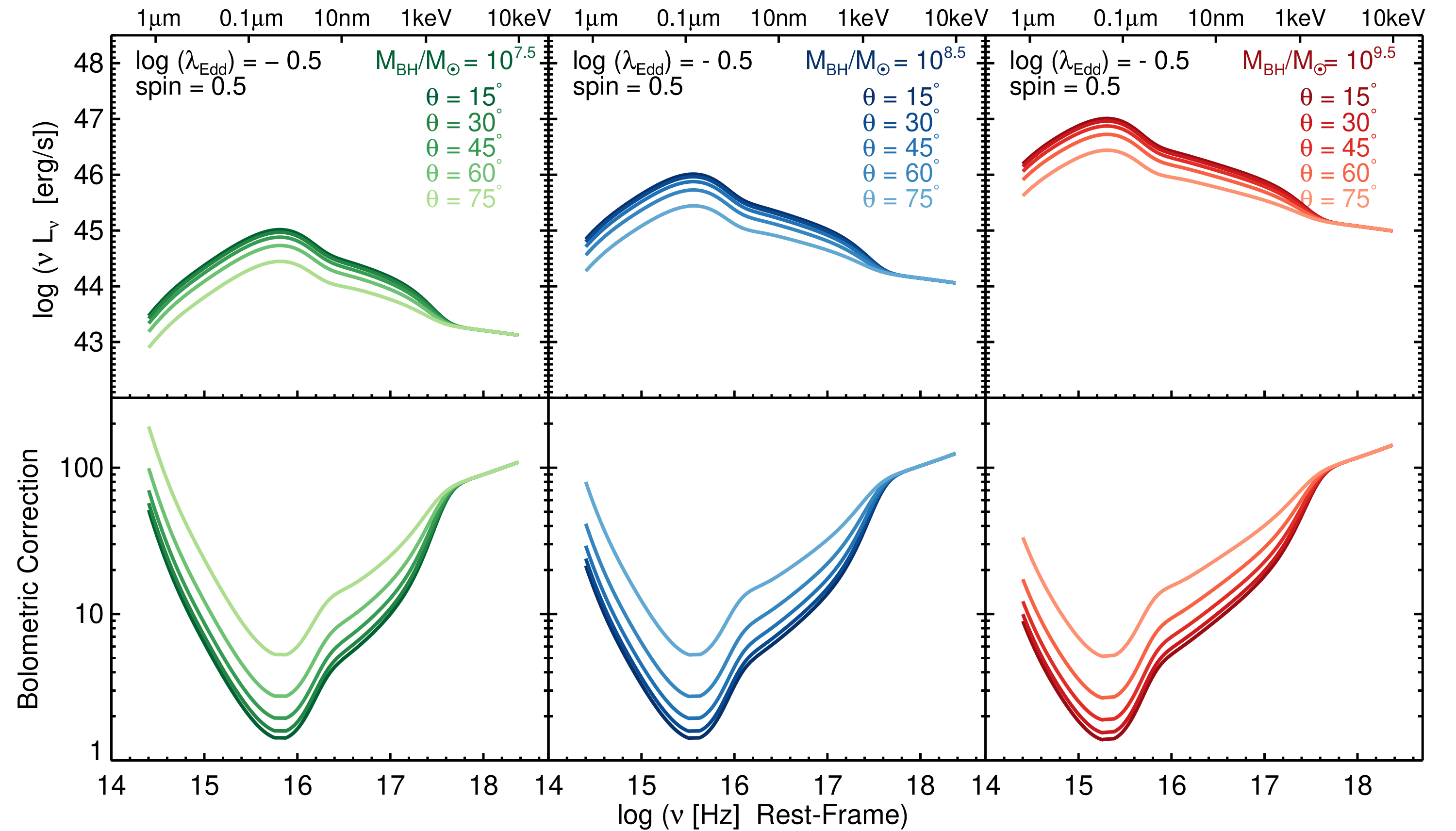}
    \caption{Accretion-disk templates (top row) generated using the QSOSED model of \citet{Kubota2018} and corresponding bolometric corrections (bottom row). The SMBH mass increases from left to right with values of $\log (M_{\mathrm{BH}}/M_\odot) = 7.5, 8.5,$ and $9.5$. The Eddington ratio is fixed at $\log \lambda_{\mathrm{Edd}} = -0.5$, and the spin is fixed at $a_\ast = 0.5$ in all panels. Within each panel, the inclination angle increases from $15^{\circ}$ (darkest color) to $75^{\circ}$ (lightest color).}
    \label{fig:th_mass_subset}
\end{figure*}

\begin{figure*}[h!]
    \centering
    \includegraphics[height=0.50\textwidth]{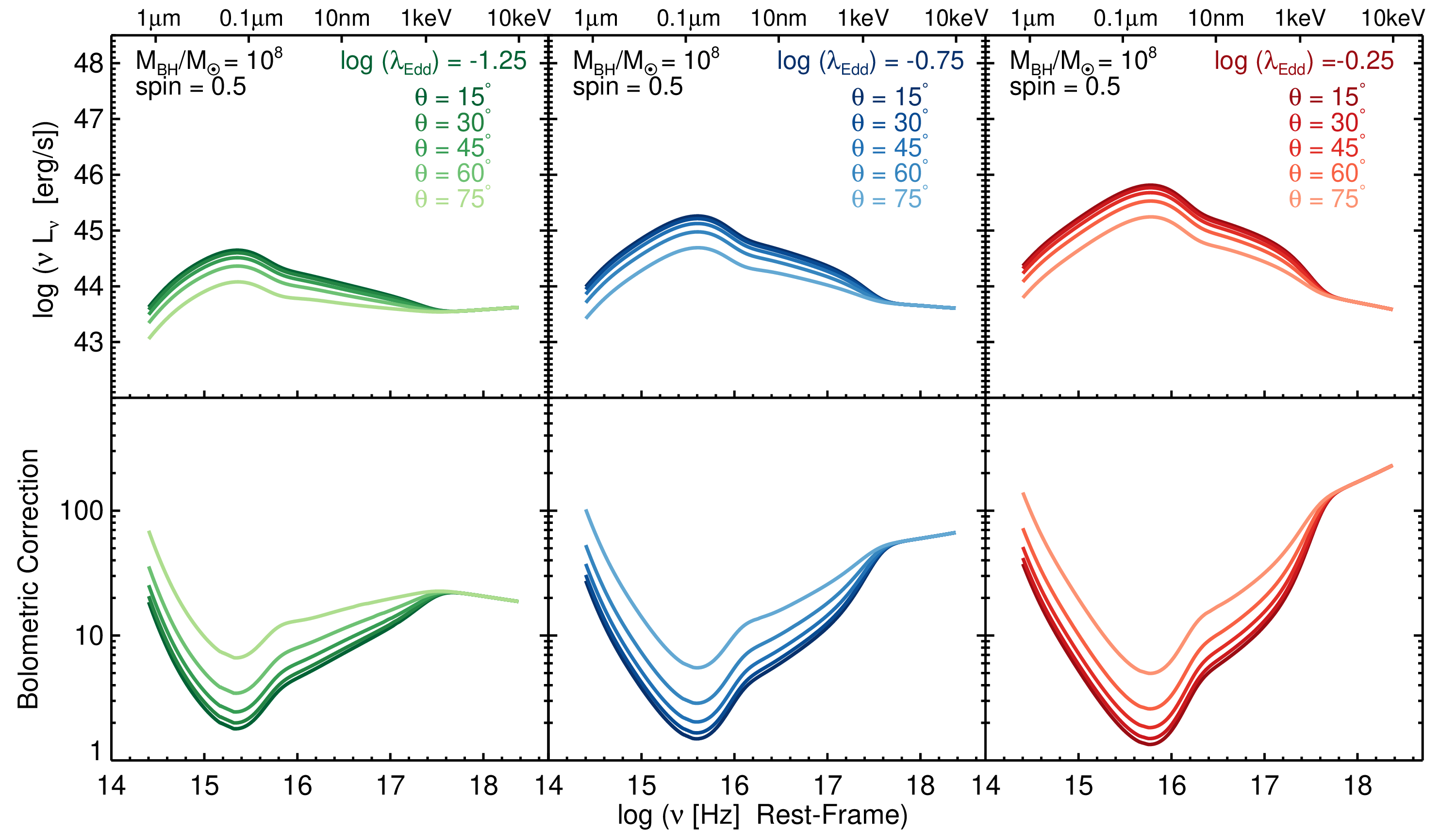}
    \caption{Accretion-disk templates (top row) generated using the QSOSED model of \citet{Kubota2018} and corresponding bolometric corrections (bottom row). The Eddington ratio increases from left to right with values of $\log \lambda_{\mathrm{Edd}}  = -1.25$, $-0.75$, and $-0.25$. The SMBH mass is fixed at $\log (M_{\mathrm{BH}}/M_\odot) = 8$, and the spin is fixed at $a_\ast = 0.5$ in all panels. Within each panel, the inclination angle increases from $15^{\circ}$ (darkest color) to $75^{\circ}$ (lightest color).}
    \label{fig:th_mdot_subset}
\end{figure*}

An accretion disk is the primary radiation source in  AGN, and its radiation dominates the visible--UV--X-ray SED in Type~1 sources. To replicate the emission from AGN accretion disks, we used the \cite{Kubota2018} QSOSED model, which has been successfully employed to describe the accretion-disk emission of both radio-quiet and radio-loud AGN populations \citep[e.g.,][]{Azadi2020, Mitchell2023, Temple2023, Kynoch2023}. QSOSED includes only the accretion disk and its immediate surroundings, not the dusty ``torus'' from which extinction arises.

QSOSED treats emission from the accretion disk as originating from three distinct regions:  1) an inner region (i.e., corona) with electron temperature $kT_{e} \sim 40$--100\,keV where hard X-ray emission ($\sim$1--10\, keV) originates; 
2) an intermediate region with electron temperature ${\sim}0.1$--1\,keV where warm Comptonization occurs and soft X-ray emission (0.01--1 keV) originates; 
and 3) an outer region where the thermal UV--visible emission originates.  The X-ray power-law originates from the innermost region, while the outer region is dominated by blackbody emission. 
At lower energies  (i.e., 0.1--2 keV), many AGNs exhibit a soft X-ray excess above the extension of the hard X-ray power-law, though its origin remains uncertain. \cite{Kubota2018} propose that this soft X-ray excess is due to the Comptonization of thermal photons by a warm ($kT_{e} \sim 0.1$--1\,keV), optically thick ($\tau \sim$10–25) layer above the disk's surface \citep[see also][]{Petrucci2018}.

\cite{Kubota2018} presented two versions of the SED models: AGNSED, the full model, and QSOSED, a simplified version in which the geometry and temperatures of the corona and the intermediate region are held fixed at typical values \citep{Kubota2018, Azadi2020}. 
The four main variables in the QSOSED model are SMBH mass, the mass accretion rate, which is parameterized by the Eddington ratio ($\lambda_{\rm Edd}$),
\begin{align}
\label{eq:edd}
  &\lambda_{\rm Edd} \equiv L_{\rm bol}/L_{\rm Edd}, \ {\rm and}
  \\
  &L_{\rm Edd} \propto M_{\rm SMBH} \quad;\nonumber
\end{align}
the dimensionless spin parameter,
\begin{align}
\label{eq:spin}
  &a\equiv c J /GM_{\rm SMBH}^{2}\quad,
\end{align}
where $J$ is the angular momentum of the SMBH; and the inclination angle $\theta$. Table~\ref{tab:qsosed} lists the main parameters and their acceptable 
ranges in the QSOSED model.

Utilizing the QSOSED model, we created 91\,500 SED templates corresponding to SMBH masses from $10^7$ to $10^{10}$~$M_{\odot}$ in steps of 0.05~dex, Eddington ratio from  0.03 to 1.0 in steps of 0.025 dex, spin from $-0.998$ to 0.998 in steps of 0.5, and inclination angle varying from 15\arcdeg\ to 75\arcdeg\ in steps of 15\arcdeg\ (Table~\ref{tab:qsosed}). Limiting the inclination angle to $\ge$15\arcdeg\ avoids complications related to beaming in blazars. Not considering $>$75\arcdeg\ avoids having to consider large and uncertain obscuration from the torus. Our analysis is restricted to a narrower black-hole mass range than the QSOSED models because we focus exclusively on SMBHs rather than including intermediate-mass black holes. We also limited the Eddington ratio to values below the Eddington limit to ensure the applicability of the thin disk model. 
The calculated SEDs are intrinsic and presented in the rest frame.

Figures~\ref{fig:new_ad_mass} and~\ref{fig:new_ad_mdot} show how the QSOSED model accretion-disk SEDs respond to changes in the Eddington ratio and SMBH mass, respectively, over a range of spin values, while holding the other parameters fixed. An increasing SMBH mass (Figure~\ref{fig:new_ad_mass}) produces a cooler and more luminous accretion disk. Increasing the Eddington ratio (Figure~\ref{fig:new_ad_mdot}) results in a more luminous and hotter disk. As spin varies from retrograde to prograde values (while other parameters are held constant), the innermost stable circular orbit ($R_{\rm ISCO}$) moves towards the SMBH, resulting in a more luminous and hotter disk with higher radiative efficiency.

The impact of the inclination angle can be seen more clearly in the top panel of Figures~\ref{fig:th_mass_subset} and~\ref{fig:th_mdot_subset}, which show only one panel from 
Figures~\ref{fig:new_ad_mass} and~\ref{fig:new_ad_mdot} but enlarged to reveal detail.  
An increase in the inclination angle (from face-on towards edge-on) results in a smaller UV bump. However, the X-ray part of the SED does not vary with the inclination angle because X-ray emission is less affected by obscuration. 

The top right panels of Figures~\ref {fig:th_mass_subset} and~\ref {fig:th_mdot_subset} also highlight the overlap between the accretion disk SEDs for different parameter values. 
 This overlap becomes particularly evident when moving from left to right in the top panels of both figures. 
For instance, an SMBH with a mass and spin fixed at \( \rm 10^8~M_{\odot} \) and 0.5, respectively, observed at an inclination angle of \( 75^\circ \) with
$\log (\lambda_{\rm Edd}) =-0.25$, exhibits a visible SED similar to that of an SMBH with $ \log (\lambda_{\rm Edd}) = -0.75$ observed at an inclination angle of \( 45^\circ \) (see the middle and right panels of Figure \ref{fig:th_mdot_subset}.) This demonstrates that the physical properties of SMBHs derived solely from SED modeling can have uncertainties, and incorporating prior information (inclination angle, spectroscopic information) is essential for constraining the fits. However, the difference of 0.5 dex in $\log \lambda_{\rm Edd}$ cited above is relatively modest, and even spectroscopic measurements often carry larger uncertainties in SMBH mass or Eddington ratio estimates. The difference in inclination angle appears more significant and, based on prior knowledge (e.g., Type 1 vs. Type 2 AGN), can be constrained with less uncertainty.

\section{Bolometric Luminosity Corrections for A General AGN population} \label{sec:bc_all}

\begin{figure*} 
    \centering
    \includegraphics[width=0.999 \textwidth,angle =0]{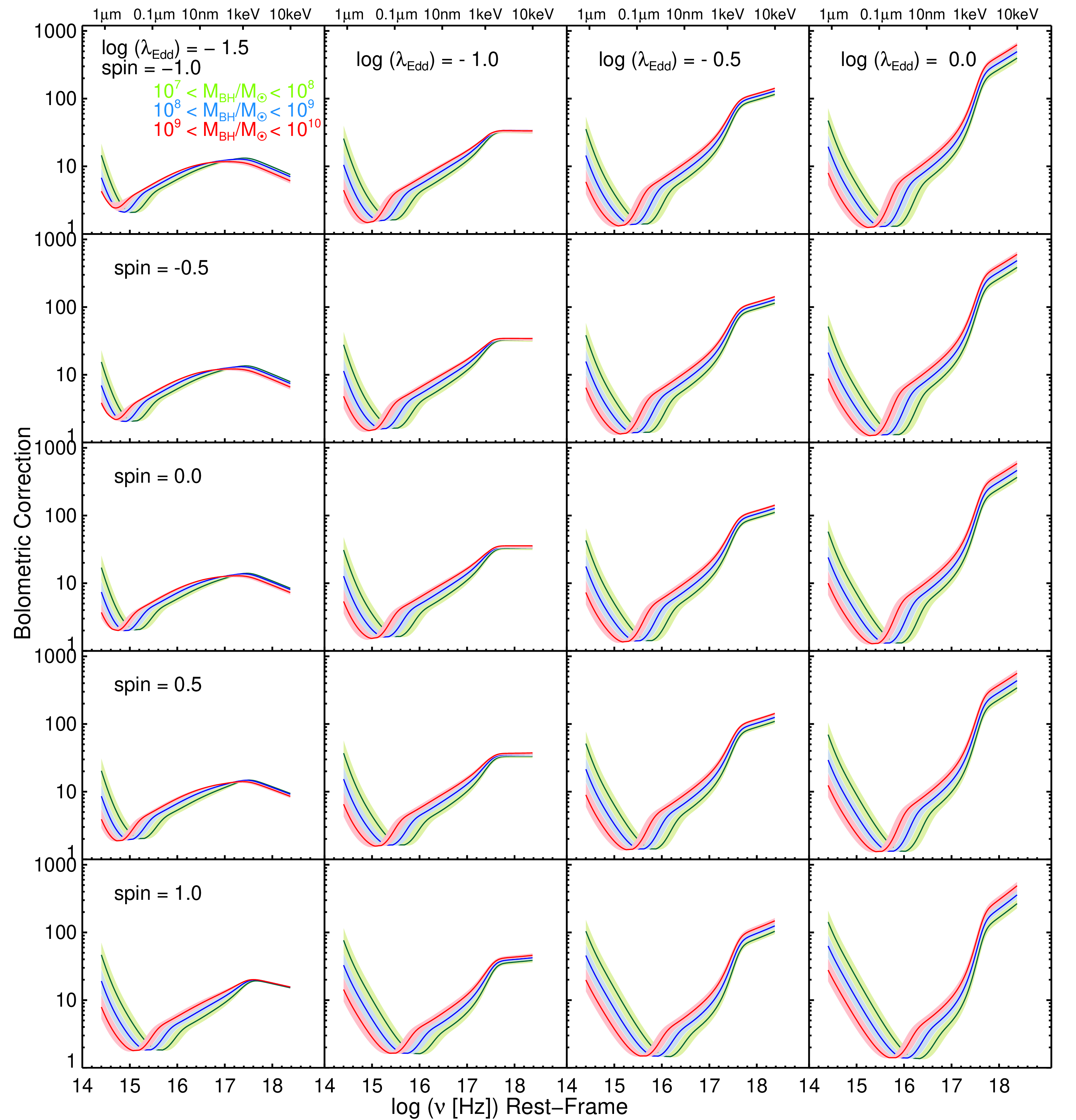}
\caption{Bolometric correction factors as a function of frequency (calculated by integrating the QSOSED templates presented in Figure~\ref{fig:new_ad_mass} from 1~\micron\ to 10~keV and over inclination angles from $0^\circ$–$90^\circ$). In each panel, the SMBH mass varies from $10^{7}$ to $10^{10}$~\Msun, with each increasing decade represented by green, blue, and red shaded areas, respectively. The Eddington ratio increases from left to right but is fixed within each panel as labeled, while the spin varies from retrograde to prograde from top to bottom. Within each panel, the inclination angle increases from $15^{\circ}$ to $75^{\circ}$; however, for clarity, only the curves corresponding to $15^{\circ}$ are shown here, with the effects of inclination illustrated in Figure~\ref{fig:th_mass_subset}. Numerical values are provided in Table \ref{tab:app3} in the appendix.} 
\label{fig:new_bc_mass}
\end{figure*}

\begin{figure*}
\begin{center}
\vspace{0cm}
\includegraphics[width=0.999\textwidth,angle =0]{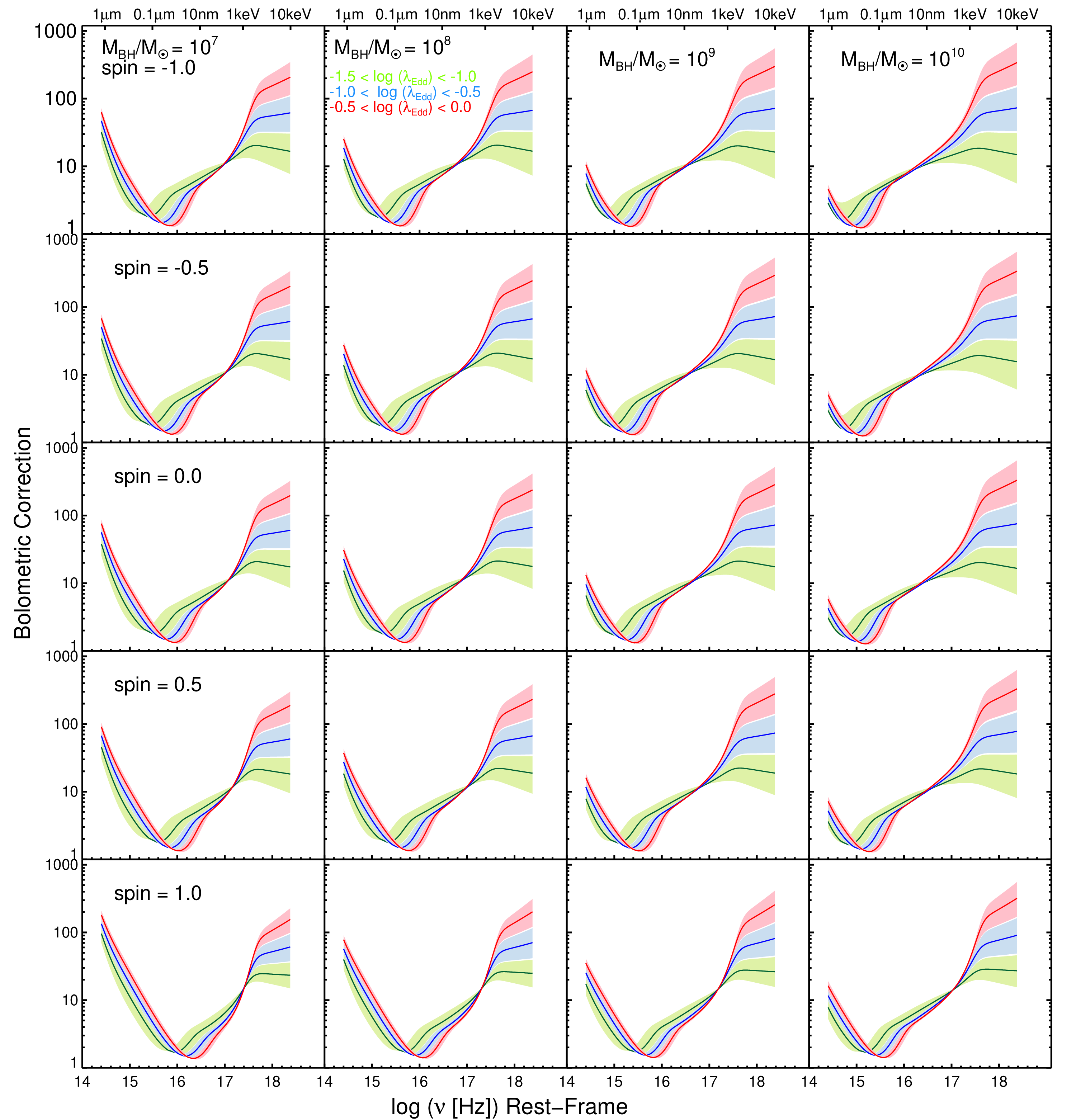}
\caption{Bolometric correction factors as a function of frequency (calculated by integrating the QSOSED templates in Figure~\ref{fig:new_ad_mdot} from 1~\micron\ to 10~keV and over inclination angles from $0^\circ$–$90^\circ$). In each panel, the Eddington ratio varies from $\log(\lambda_{\rm Edd})=-1.5$ to $0.0$, with each increasing decade represented by green, blue, and red shaded areas, respectively. The SMBH mass increases from left to right but is fixed within each panel as labeled, while the spin varies from retrograde to prograde from top to bottom. Within each panel, the inclination angle increases from $15^{\circ}$ to $75^{\circ}$; however, for clarity, only the curves corresponding to $15^{\circ}$ are shown here, with the effects of inclination illustrated in Figure~\ref{fig:th_mdot_subset}. Numerical values are provided in Table \ref{tab:app4} in the appendix.}   
\label{fig:new_bc_mdot}
\end{center}
\end{figure*}

To estimate the bolometric luminosity, we integrated the intrinsic accretion-disk templates introduced in Section~\ref{sec:ad} from 1\,\um to 10\,keV. The bolometric correction factor for a given $M_{\rm BH}$, $\lambda_{\rm Edd}$, and spin is then defined as
\begin{equation} \label{eq:bc}
    {\rm BC}({\nu},\theta) \equiv \frac{\int _0^{\pi/2 }\sin(\theta)\int _{1\,\mu{\rm m}}^{10\,\rm keV} L_{\nu}(\theta) d{\nu}d\theta}{\nu L_{\nu}(\theta)} 
\end{equation}
for any specified frequency $\nu$ and inclination angle $\theta$.
MIR emission from the torus is not included in this integral because that emission is dominated by reprocessed accretion-disk radiation, which is already included. The accretion-disk templates self-consistently cover the usual gap in observations between the UV and soft X-ray bands, and therefore, no gap repair is required. Additionally, the model yields both the intrinsic luminosity and the observed SED, accounting for the effects of the inclination angle (as well as SMBH mass, Eddington ratio, and spin). Therefore, there is no need to apply any further correction for the anisotropy of the accretion-disk emission. 
Figures~\ref{fig:new_bc_mass} and~\ref{fig:new_bc_mdot} show the calculated bolometric corrections as a function of frequency. The input parameters have complex, interrelated effects on both the accretion-disk SED and the bolometric correction. Obtaining independent estimates of one or more of these parameters will result in more accurate bolometric-correction estimates. We discuss the impact of variations of each of the parameters on the bolometric correction function below.

\subsection{SMBH Mass Dependence} \label{sec:mass}

Figure ~\ref{fig:new_bc_mass} (based on the SEDs presented in  Figure~\ref{fig:new_ad_mass}) shows how the bolometric correction responds to variations in SMBH mass with the other parameters fixed. 
The dip in the bolometric correction at $10^{15}$--$10^{16}$\,Hz corresponds to the peak in the accretion disk SED 
(Figure~\ref{fig:new_ad_mass}). As SMBH mass increases, 
this dip shifts to lower frequencies, corresponding to a cooler accretion disk.
At frequencies lower than the dip, more massive SMBHs require a smaller bolometric correction than their lower-mass counterparts. This is because the lower frequencies are closer to the peak of the accretion disk emission for more massive SMBHs. 
The opposite happens at frequencies greater than the dip: smaller bolometric corrections are needed for lower SMBH masses because the peak of accretion disk emission occurs at higher frequencies. 
The SMBH mass has the least impact on the bolometric correction near the dip in the SED and at X-ray frequencies. At X-rays, however, at higher Eddington ratios, the dependence on mass becomes slightly more pronounced. In other words, the fraction of the bolometric luminosity emitted near the peak of the accretion disk SED and at X-rays is nearly independent of SMBH mass but has significant dependence on Eddington ratio (see Section \ref{sec:ledd}).

\subsection{Eddington-Ratio Dependence} \label{sec:ledd}
Figure~\ref{fig:new_bc_mdot} (based on the SEDs presented in Figure~\ref{fig:new_ad_mdot}) shows how the bolometric correction responds to variations in Eddington ratio with the other parameters fixed. As the Eddington ratio increases, the minimum bolometric correction, corresponding to the peak of the accretion disk emission, occurs at higher frequencies. At frequencies below the dip, AGN with lower Eddington ratios require a smaller correction than their higher Eddington-ratio counterparts because the peak of their accretion disk occurs at lower frequencies. In contrast, the opposite happens at frequencies above the dip.

In the X-ray band, the accretion-disk templates converge for the same SMBH mass and spin
(Figure~\ref{fig:new_ad_mdot})
to give the same $\nu L_{\nu}$ independent of Eddington ratio. This is inherent in the QSOSED model and is due to a combination of factors: an increase in Eddington ratio results in increased luminosity \citep{Ho2008} and a steepening of $\alpha_{ox}$ and $\Gamma$(2--10 keV) (For more details see \citealt{Kubota2018}). The combination of the convergence in the SEDs and the difference in the integrated luminosity 
for AGN of different Eddington ratios results in a large range of bolometric correction in the X-ray regime (Figure~\ref{fig:new_bc_mdot}).

\subsection{Spin Dependence}

Figures~\ref{fig:new_bc_mass} and~\ref{fig:new_bc_mdot} also reveal how varying SMBH spin affects the bolometric correction. Varying the spin from retrograde to prograde (with all other parameters fixed) shifts the accretion-disk SED slightly toward higher energies, producing a hotter and more luminous disk. Therefore, the dip in the bolometric correction function (corresponding to the peak of accretion disk SED) moves towards higher frequencies. 
However, spin has relatively little effect on the bolometric correction except that the largest prograde spin makes the correction slightly higher. 

The spin values considered here cover the broadest possible range of AGN. However, as shown in Figure~\ref{fig:new_ad_mass} and \ref{fig:new_ad_mdot}, small changes in spin do not significantly change the accretion disk SED, resulting in a weak dependence of the bolometric correction on SMBH spin in a given sample (e.g., radio-loud or radio-quiet) (see Figure \ref{fig:bc_mass_wave_spin} and Section \ref{sec:discussion} for more discussion).


\subsection{Inclination Angle Dependence} \label{sec:th}

Figures~\ref{fig:th_mass_subset} and~\ref{fig:th_mdot_subset} show the impact of the inclination angle on the SED and bolometric correction. In all cases, a face-on observer ($15^{\circ}$) sees a more luminous accretion disk in the visible–-UV bands that diverges more from those of an edge-on observer ($75^{\circ}$). The combination of a lower integrated luminosity and a lower \(\nu L_{\nu}\) value results in a slightly higher bolometric correction for the edge-on case. 
At X-ray energies, this effect is less pronounced, and the difference between the templates is smaller than at lower frequencies. This is because the X-rays, in the QSOSED model, emit isotropically. 

In summary, Figures~\ref{fig:new_ad_mass} to~\ref{fig:new_bc_mdot} show how SMBH mass, Eddington ratio, spin, and inclination angle change the accretion-disk SED
and bolometric correction. An observer with photometry---corrected for extinction---in the range from 1\,\um to 10\,keV can apply the SEDs presented here
to replicate the accretion disk emission of radio-silent
to radio-loud AGN at any redshift. Then, the bolometric
power of these AGN can be estimated
using the bolometric correction factors from the figures above. Tables in the Appendix give the SEDs and the bolometric correction values.


\begin{figure*}[th!]
\begin{center}
    \includegraphics[width=\textwidth,angle =0]{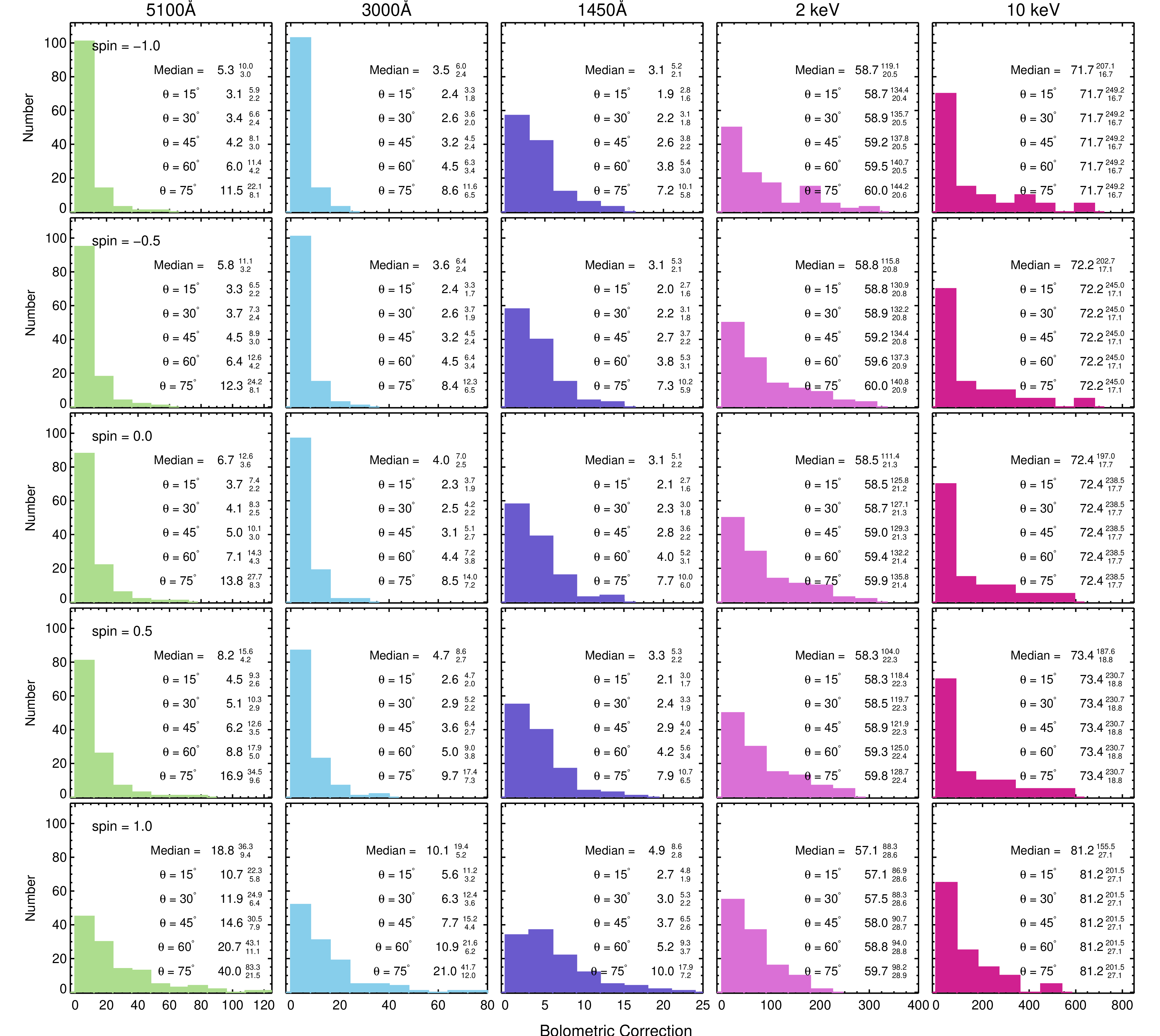}
 \caption{The distributions of bolometric corrections at 5100\,\AA, 3000\,\AA, 1450\,\AA, 2\,keV, and 10\,keV for retrograde ($\rm spin = -1.0$) to prograde ($\rm spin = 1.0$) SMBHs. Distributions are based on the median bolometric correction curves in Figures \ref{fig:new_bc_mass} and \ref{fig:new_bc_mdot}, thus depend on SMBH mass, Eddington ratio, spin, and inclination angle. The parameter space covered includes SMBH masses in the range $10^{7}$--$10^{8}\,M_\odot$, Eddington ratios spanning $\log \lambda_{\rm Edd} = -1.5$ to $0.0$, and the inclination angles and spin values highlighted in the figure. The median values for each distribution and the corresponding 25th--75th percentile envelopes are given in each panel. Additionally, the median value corresponding to each inclination angle is also presented.}    
\label{fig:bc_hist}
\end{center}
\end{figure*}

\section{Discussion} \label{sec:discussion}

In this section, we first estimate the bolometric corrections using our technique at five commonly used values in the literature (5100\,\AA, 3000\,\AA, 1450\,\AA, 2\,keV, and 10\,keV) and investigate how these values depend on the AGN properties. 
We then compare our findings with the literature. 
For the analysis in this section, we used the median bolometric correction curves presented in Figures~\ref{fig:new_bc_mass} and \ref{fig:new_bc_mdot}. While only the $15^{\circ}$ inclination curve is displayed in those figures for clarity, we incorporated the corresponding values for other inclination angles ($30^{\circ}$ to $75^{\circ}$) in our analysis. Therefore, our bolometric correction estimates reflect variations in SMBH mass, Eddington ratio, spin, and inclination angle.

\subsection{AGN Bolometric Corrections Across Key Wavelengths and Parameters}  \label{sec:main_plot}

Figure~\ref{fig:bc_hist} presents the distributions of bolometric corrections across different wavebands. For any combination of parameters, the bolometric correction generally decreases from the visible to the FUV bands but is highest in the X-ray (Figure~\ref{fig:bc_hist}).
The bolometric correction is smallest at 1450\,\AA, making it the most efficient single-band estimator of the AGN bolometric luminosity. This is because 1450\,\AA\ lies closest to the peak of the accretion disk SED\null.
These trends are verified by the median values shown in Figure~\ref{fig:bc_hist}.
At the two longest wavelengths, as the spin increases from $0.5$ to $+1.0$, the median bolometric correction shows an increase. The effect of spin is smaller in the far-UV and almost negligible in X-rays. As noted in Section~\ref{sec:bc_all}, this trend arises because the innermost stable circular orbit lies closer to the SMBH for $\rm spin = +1$, producing a hotter and more luminous accretion disk. As a result, the visible portion of the SED accounts for a smaller fraction of the total radiative output; consequently, a higher bolometric correction is needed. However, this spin dependence vanishes at X-ray energies. Indeed, a Kolmogorov--Smirnov (K--S) test\footnote{The K--S probabilities represent the likelihood that two samples drawn from the same distribution would differ as much as observed.} comparing the distributions for $\rm spin = -1$ and $\rm spin = +1$ yields probabilities of $2.3 \times 10^{-14}$, $2.4 \times 10^{-11}$, $8.5 \times 10^{-4}$, $6.3 \times 10^{-2}$, and $6.3 \times 10^{-2}$ at 5100\,\AA, 3000\,\AA, 1450\,\AA, 2\,keV, and 10\,keV, respectively. These results indicate highly significant differences at visible and UV wavelengths, while at X-ray energies, the distributions are statistically consistent with being drawn from the same parent population.

Each panel of Figure~\ref{fig:bc_hist} also shows the median bolometric correction at each inclination angle, progressing from a nearly face-on view ($15^\circ$) to an edge-on perspective ($75^\circ$). In the visible–UV bands, an edge-on view tends to require higher bolometric corrections than face-on observers, though the distributions overlap. In contrast, at X-ray energies, the inclination dependence largely disappears, with all viewing angles having essentially the same bolometric correction. This is because, as shown in Figures~\ref{fig:th_mass_subset} and~\ref{fig:th_mdot_subset}, the X-ray SEDs overlap closely, leading to similar bolometric corrections regardless of inclination. Overall, Figure~\ref{fig:bc_hist} demonstrates that, at X-ray energies, the bolometric correction is effectively independent of both spin and inclination angle.

\begin{figure*}[!ht]
\begin{center}
    \includegraphics[height=0.9\textwidth,angle =0]{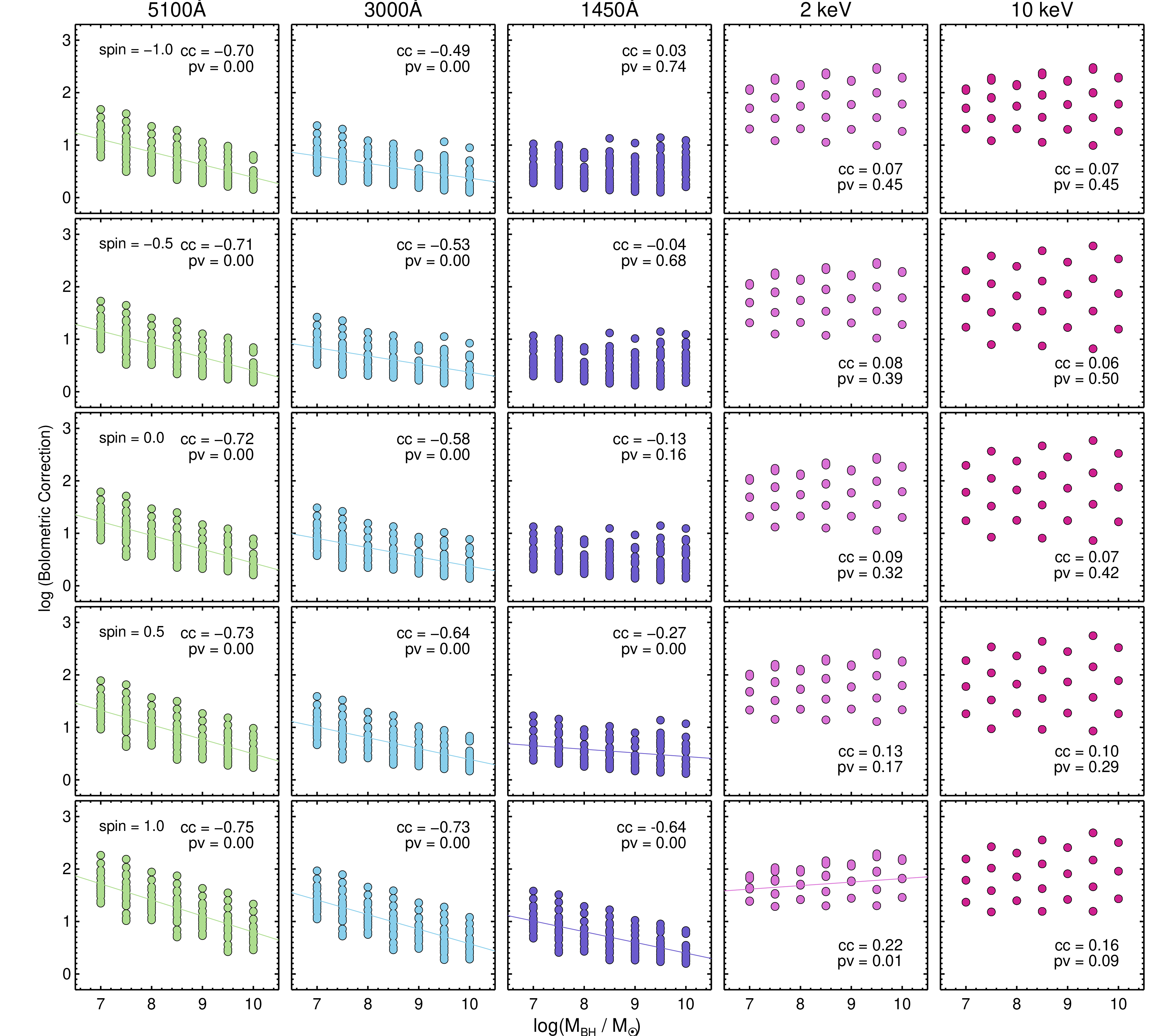}
 \caption{Relation between bolometric correction and SMBH mass at 5100\,\AA, 3000\,\AA, 1450\,\AA, 2\,keV and 10\,keV for retrograde ($\mathrm{spin} = -1.0$) to prograde ($\mathrm{spin} = 1.0$) SMBHs. The data points are derived from the median bolometric correction curves in Figures~\ref{fig:new_bc_mass} and~\ref{fig:new_bc_mdot} and therefore depend on SMBH mass, Eddington ratio, spin, and inclination angle. The correlation coefficients and their significances are shown in each panel. For cases with significance ${>}3\sigma$, solid lines show the best linear fit, and the best-fit parameters are reported in Table~\ref{tab:eq_mass}. }    
\label{fig:bc_mass_wave_spin}
\end{center}
\end{figure*}

\begin{figure*}[!ht]
\begin{center}
    \includegraphics[height=0.9\textwidth,angle =0]{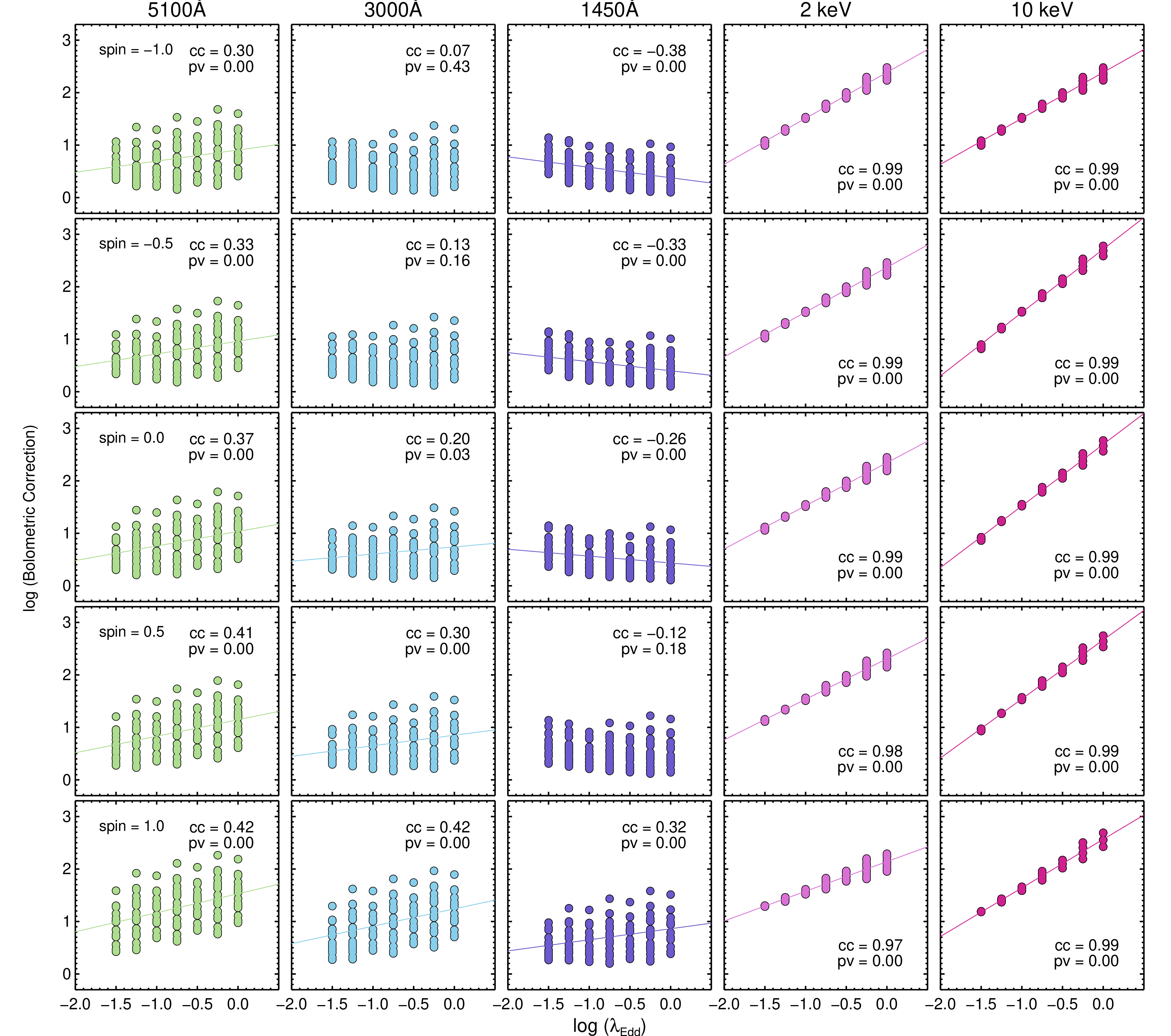}
 \caption{Relation between bolometric correction and the Eddington ratio at 5100\,\AA, 3000\,\AA, 1450\,\AA, and 10\,keV for retrograde ($\mathrm{spin} = -1.0$) to prograde ($\mathrm{spin} = 1.0$) SMBHs. The data points are derived from the median bolometric correction curves in Figures~\ref{fig:new_bc_mass} and~\ref{fig:new_bc_mdot}, and therefore depend on SMBH mass, Eddington ratio, spin, and inclination angle. The correlation coefficients and their significances are shown in each panel. For cases with significance $>3\sigma$, a linear fit is applied, and the best-fit parameters are reported in Table~\ref{tab:eq_mdot}. }    
\label{fig:bc_mdot_wave_spin}
\end{center}
\end{figure*} 

\begin{figure*}[!ht]
\begin{center}
    \includegraphics[height=0.9\textwidth,angle =0]{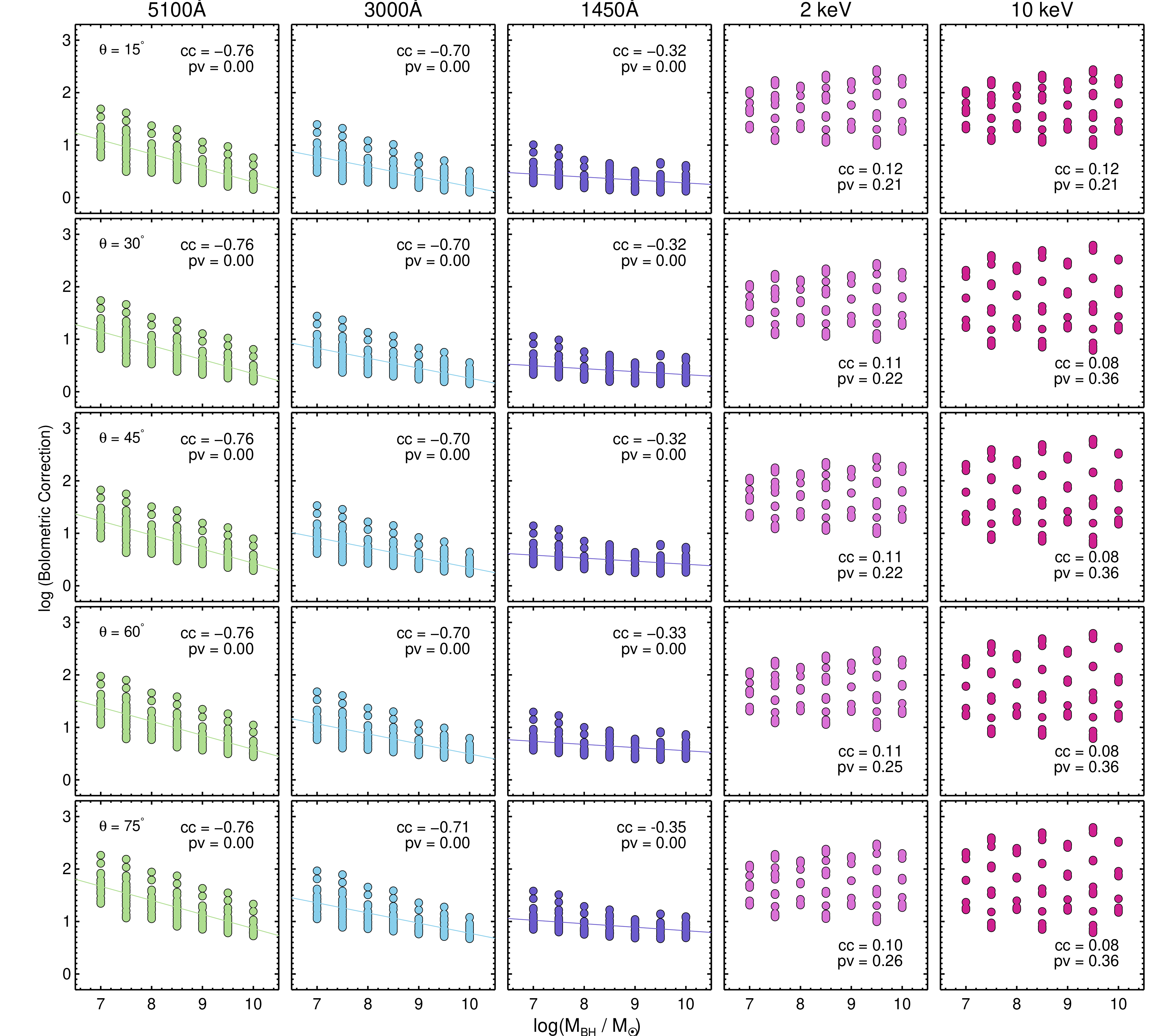}
 \caption{Relation between bolometric correction and SMBH mass at 5100\,\AA, 3000\,\AA, 1450\,\AA, and 10\,keV for different black-hole masses. The data points are derived from the median bolometric correction curves in Figures~\ref{fig:new_bc_mass} and~\ref{fig:new_bc_mdot}, and therefore depend on SMBH mass, Eddington ratio, spin, and inclination angle. The correlation coefficients and their significances are shown in each panel. For cases with significance $>3\sigma$, a linear fit is applied, and the best-fit parameters are reported in Table~\ref{tab:eq_mass_th}. }    
\label{fig:bc_mass_wave_th}
\end{center}
\end{figure*}

\begin{figure*}[!ht]
\begin{center}
    \includegraphics[height=0.9\textwidth,angle =0]{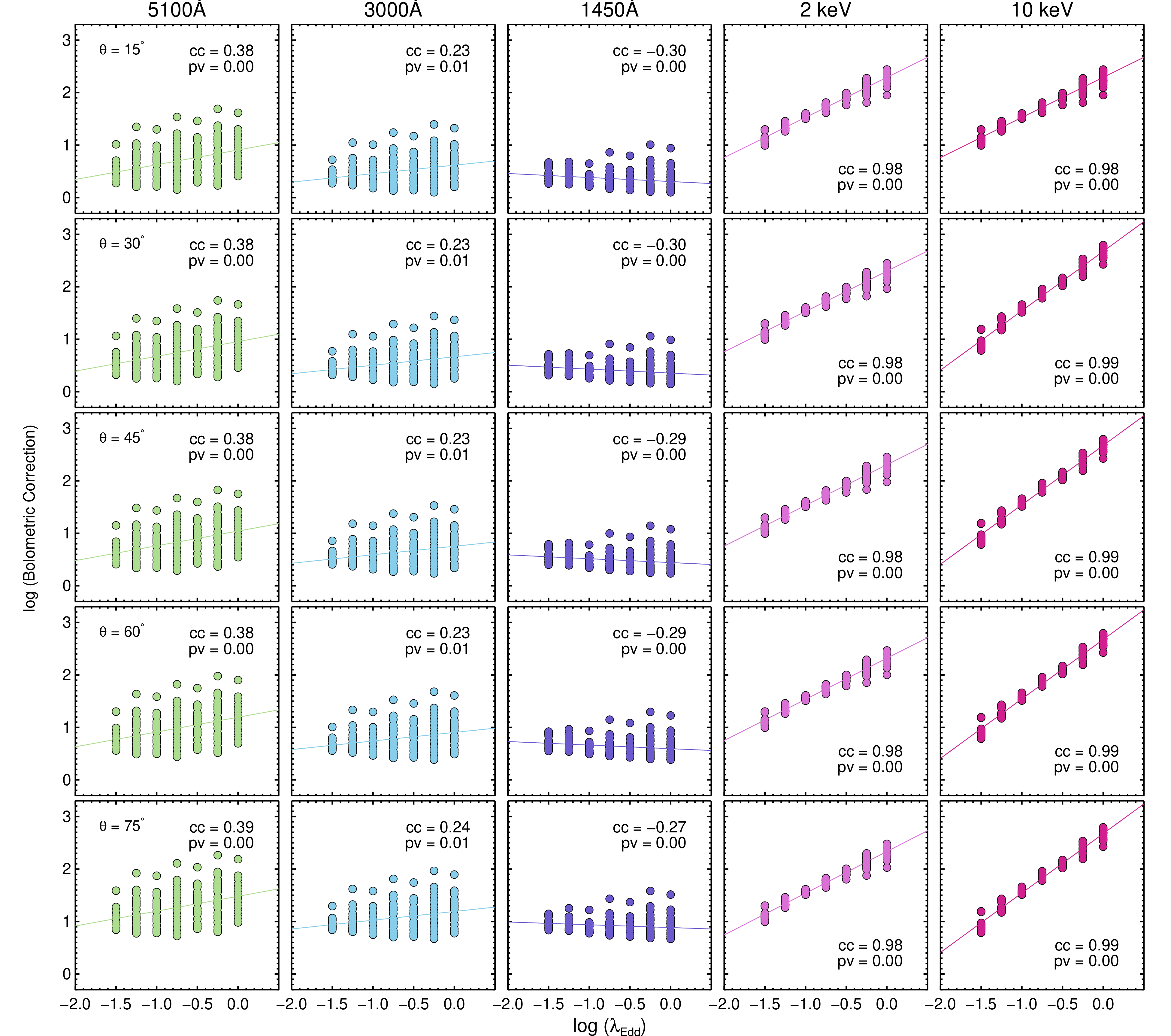}
 \caption{Relation between bolometric correction and the Eddington ratio at 5100\,\AA, 3000\,\AA, 1450\,\AA, and 10\,keV for different inclination angles. The data points are derived from the median bolometric correction curves in Figures~\ref{fig:new_bc_mass} and~\ref{fig:new_bc_mdot}, and therefore depend on SMBH mass, Eddington ratio, spin, and inclination angle. The correlation coefficients and their significances are shown in each panel. For cases with significance $>3\sigma$, a linear fit is applied, and the best-fit parameters are reported in Table~\ref{tab:eq_mdot_th}. }    
\label{fig:bc_mdot_wave_th}
\end{center}
\end{figure*}

\subsection{Dependence of Bolometric Corrections on Black Hole Mass and Accretion Rate} \label{sec:wavelengths}

Figures~\ref{fig:bc_mass_wave_spin} and~\ref{fig:bc_mdot_wave_spin} show how the bolometric correction varies as a function of SMBH mass and Eddington ratio, respectively, for different spin values. As noted in Figure~\ref{fig:bc_hist}, we used the median bolometric correction curves from Figures~\ref{fig:new_bc_mass} and~\ref{fig:new_bc_mdot} to derive the data points.
To quantify the relationship between bolometric correction and either SMBH mass or Eddington ratio, we computed the Spearman rank correlation coefficient (cc) and its significance (pv) using IDL’s $r-correlate $ routine. For cases where the significance exceeds $>3\sigma$, we fitted a linear relation; the best-fit parameters are listed in Tables~\ref{tab:eq_mass} and~\ref{tab:eq_mdot}.


Figure~\ref{fig:bc_mass_wave_spin} shows that, at visible and NUV wavelengths, the bolometric correction generally decreases with increasing $M_{\rm SMBH}$; this trend weakens at 1450\,\AA\ and is not detected in the X-rays (see Table~\ref{tab:eq_mass}). 
This behavior is expected: as discussed in Section~\ref{sec:mass}, increasing SMBH mass cools the disk and shifts its thermal peak to longer wavelengths, naturally producing a stronger mass dependence at 5100 and 3000\,\AA\ than at 1450\,\AA. 
At 1450\,\AA, a weak but statistically significant dependence appears only for spin $>0$, when the disk peak moves to shorter wavelengths. 
In the X-rays, the dependence disappears for almost all spins, consistent with the hard X-ray emission originating primarily in the corona and being largely independent of SMBH mass and spin.

Figure \ref{fig:bc_mdot_wave_spin} reveals a large variation in the relation between the bolometric correction and the Eddington ratio.
At 5100\,\AA\ and 3000\,\AA, the bolometric correction increases with Eddington ratio for all spin values, with the slope slightly steepening as spin increases.  At 1450\,\AA, the correlation shifts from negative to positive as spin increases, indicating a strong spin dependence at this wavelength. At X-rays, the bolometric correction exhibits a consistently strong, positive correlation with Eddington ratio across all spins, reflecting the sensitivity of 2--10 keV emission to the Eddington ratio (Table~\ref{tab:eq_mdot}).


Figures~\ref{fig:bc_mass_wave_th} and~\ref{fig:bc_mdot_wave_th} show how bolometric corrections vary with SMBH mass and Eddington ratio as a function of inclination. In Figure~\ref{fig:bc_mass_wave_th}, the bolometric correction–$M_{\rm SMBH}$ relation is negative at visible--NUV wavelengths, flattens toward 1450\,\AA, and vanishes in the X-rays. As summarized in Table~\ref{tab:eq_mass_th}, this trend shows no significant dependence on inclination at any wavelength. By contrast, Figure~\ref{fig:bc_mdot_wave_th} reveals stronger band-to-band differences in the slope of the bolometric correction–Eddington ratio relation; the slope varies only slightly with inclination (Table~\ref{tab:eq_mdot_th}).


Overall, Figures~\ref{fig:bc_mass_wave_spin}--\ref{fig:bc_mdot_wave_th} demonstrate a strong dependence of the bolometric correction on the Eddington ratio, most prominently in the X-rays, and a milder dependence on SMBH mass at visible–NUV wavelengths (as seen by comparing the slopes of the relations in Tables~\ref{tab:eq_mass} to \ref{tab:eq_mdot_th}). The strength of these dependencies shows moderate variation with spin and only weak variation with inclination angle. In contrast, the FUV bands exhibit the weakest dependence on the fundamental AGN parameters, making these wavelengths an especially robust regime for estimating bolometric luminosity.

\begin{table*}
\centering
\caption{Best-fit parameters for BC as a function of $M_{\rm  SMBH}$ from Figure \ref{fig:bc_mass_wave_spin}}
\resizebox{\textwidth}{!}{%
\begin{tabular}{lccccc}
    \hline \hline
        & 5100\,\AA & 3000\,\AA & 1450\,\AA & 2\,keV & 10\,keV \\
        & $m\pm\Delta m$, $c\pm\Delta c$ & $m\pm\Delta m$, $c\pm\Delta c$ & $m\pm\Delta m$, $c\pm\Delta c$ & $m\pm\Delta m$, $c\pm\Delta c$ & $m\pm\Delta m$, $c\pm\Delta c$ \\
    \hline 
spin = $-1.0$ & $-0.24\pm0.02$, $2.80\pm0.20$ & $-0.14\pm0.02$, $1.80\pm0.19$ & \no & \no & \no \\
spin = $-0.5$ & $-0.25\pm0.02$, $2.90\pm0.20$ & $-0.16\pm0.02$, $1.90\pm0.19$ & \no & \no & \no \\
spin = $0.0$  & $-0.26\pm0.02$, $3.10\pm0.20$ & $-0.18\pm0.02$, $2.10\pm0.19$ & \no & \no & \no \\
spin = $0.5$  & $-0.28\pm0.03$, $3.30\pm0.21$ & $-0.21\pm0.02$, $2.50\pm0.20$ & $-0.07\pm0.02$, $1.10\pm0.20$ & \no & \no \\
spin = $1.0$  & $-0.31\pm0.03$, $3.90\pm0.22$ & $-0.28\pm0.03$, $3.40\pm0.21$ & $-0.20\pm0.02$, $2.40\pm0.19$ & $0.07\pm0.03$, $1.10\pm0.23$ & \no \\
\hline
\end{tabular}%
}
\label{tab:eq_mass}
\raggedright
\tablecomments{Tabulated numbers are coefficients in $\log({\rm BC}) = (m\pm \Delta m)\,\log \!\left(\frac{M_{\rm SMBH}}{\Msol}\right) + (c\pm\Delta c)$ for different spin values, averaged over inclination. Values are shown for wavelength--spin combinations with significant correlation ($p<0.05$).}
\end{table*}


\begin{table*}
\centering
\caption{Best-fit parameters for BC as a function of $\lambda_{\rm Edd}$ from Figure \ref{fig:bc_mdot_wave_spin}}
\resizebox{\textwidth}{!}{%
\begin{tabular}{cccccc}
    \hline \hline
        & 5100\,\AA & 3000\,\AA & 1450\,\AA & 2\,keV & 10\,keV \\
        & $m\pm\Delta m$, $c\pm\Delta c$ & $m\pm\Delta m$, $c\pm\Delta c$ & $m\pm\Delta m$, $c\pm\Delta c$ & $m\pm\Delta m$, $c\pm\Delta c$ & $m\pm\Delta m$, $c\pm\Delta c$ \\
    \hline
spin = $-1.0$ & $0.21\pm0.06$, $0.91\pm0.06$ & \no & $-0.20\pm0.04$, $0.38\pm0.04$ & $0.88\pm0.01$, $2.40\pm0.01$ & $0.88\pm0.01$, $2.40\pm0.01$ \\
spin = $-0.5$ & $0.24\pm0.06$, $0.96\pm0.06$ & \no & $-0.17\pm0.05$, $0.40\pm0.04$ & $0.85\pm0.01$, $2.40\pm0.01$ & $1.20\pm0.01$, $2.70\pm0.01$ \\
spin = $0.0$  & $0.28\pm0.06$, $1.00\pm0.06$ & $0.14\pm0.05$, $0.74\pm0.05$ & $-0.13\pm0.05$, $0.43\pm0.04$ & $0.82\pm0.01$, $2.30\pm0.01$ & $1.20\pm0.01$, $2.70\pm0.01$ \\
spin = $0.5$  & $0.32\pm0.07$, $1.10\pm0.06$ & $0.20\pm0.06$, $0.85\pm0.05$ & \no & $0.77\pm0.01$, $2.30\pm0.01$ & $1.10\pm0.01$, $2.70\pm0.01$ \\
spin = $1.0$  & $0.36\pm0.07$, $1.50\pm0.06$ & $0.33\pm0.07$, $1.20\pm0.06$ & $0.21\pm0.06$, $0.86\pm0.05$ & $0.57\pm0.02$, $2.10\pm0.01$ & $0.93\pm0.01$, $2.60\pm0.01$ \\
\hline
\end{tabular}%
}
\label{tab:eq_mdot}
\raggedright
\tablecomments{Tabulated numbers are coefficients in the equation $\log({\rm BC}) = (m\pm \Delta m)\,\log(\lambda_{\rm Edd}) + (c\pm\Delta c)$ for different spin values, averaged over inclination. Values are given for wavelength--spin combinations with significant correlation, $p<0.05$.}
\end{table*}

\begin{table*}
\centering
\caption{Best-fit parameters for BC as a function of $M_{\rm SMBH}$ from Figure \ref{fig:bc_mass_wave_th}}
\resizebox{\textwidth}{!}{%
\begin{tabular}{cccccc}
    \hline \hline
       & 5100\,\AA & 3000\,\AA & 1450\,\AA & 2\,keV & 10\,keV \\
       & $m\pm\Delta m$, $c\pm\Delta c$ & $m\pm\Delta m$, $c\pm\Delta c$ & $m\pm\Delta m$, $c\pm\Delta c$ & $m\pm\Delta m$, $c\pm\Delta c$ & $m\pm\Delta m$, $c\pm\Delta c$ \\
    \hline 
$\rm \theta = 15^{\circ}$ & $-0.27\pm0.02$, $3.00\pm0.19$ & $-0.19\pm0.02$, $2.10\pm0.17$ & $-0.06\pm0.02$, $0.84\pm0.14$ & \no & \no \\
$\rm \theta = 30^{\circ}$ & $-0.27\pm0.02$, $3.00\pm0.19$ & $-0.19\pm0.02$, $2.20\pm0.17$ & $-0.06\pm0.02$, $0.90\pm0.14$ & \no & \no \\
$\rm \theta = 45^{\circ}$ & $-0.27\pm0.02$, $3.10\pm0.19$ & $-0.19\pm0.02$, $2.30\pm0.17$ & $-0.06\pm0.02$, $0.99\pm0.14$ & \no & \no \\
$\rm \theta = 60^{\circ}$ & $-0.27\pm0.02$, $3.30\pm0.19$ & $-0.19\pm0.02$, $2.40\pm0.17$ & $-0.06\pm0.02$, $1.20\pm0.14$ & \no & \no \\
$\rm \theta = 75^{\circ}$ & $-0.27\pm0.02$, $3.50\pm0.19$ & $-0.19\pm0.02$, $2.70\pm0.17$ & $-0.07\pm0.02$, $1.50\pm0.14$ & \no & \no \\
\hline
\end{tabular}%
}
\label{tab:eq_mass_th}
\raggedright
\tablecomments{Tabulated numbers are coefficients in $\log({\rm BC}) = (m\pm \Delta m)\,\log \!\left(\frac{M_{\rm SMBH}}{\Msol}\right) + (c\pm\Delta c)$  for different \emph{inclination} values (averaged over spin). Values are shown for wavelength--spin combinations with significant correlation ($p<0.05$).}
\end{table*}


\begin{table*}
\centering
\caption{Best-fit parameters for BC as a function of $\lambda_{\rm Edd}$ from Figure \ref{fig:bc_mdot_wave_th}} 
\resizebox{\textwidth}{!}{%
\begin{tabular}{cccccc}
\hline\hline
       & 5100\,\AA & 3000\,\AA & 1450\,\AA & 2\,keV & 10\,keV \\
       & $m\pm\Delta m$, $c\pm\Delta c$ & $m\pm\Delta m$, $c\pm\Delta c$ & $m\pm\Delta m$, $c\pm\Delta c$ & $m\pm\Delta m$, $c\pm\Delta c$ & $m\pm\Delta m$, $c\pm\Delta c$ \\
\hline\hline
$\rm \theta = 15^{\circ}$ & $0.28\pm0.06$, $0.91\pm0.06$ & $0.16\pm0.05$, $0.62\pm0.05$ & $-0.08\pm0.04$, $0.31\pm0.03$ & $0.76\pm0.02$, $2.30\pm0.01$ & $0.76\pm0.02$, $2.30\pm0.01$ \\
$\rm \theta = 30^{\circ}$ & $0.28\pm0.06$, $0.96\pm0.06$ & $0.16\pm0.05$, $0.67\pm0.05$ & $-0.08\pm0.04$, $0.35\pm0.03$ & $0.77\pm0.02$, $2.30\pm0.01$ & $1.10\pm0.02$, $2.70\pm0.01$ \\
$\rm \theta = 45^{\circ}$ & $0.28\pm0.06$, $1.00\pm0.06$ & $0.16\pm0.05$, $0.75\pm0.05$ & $-0.07\pm0.04$, $0.44\pm0.03$ & $0.78\pm0.02$, $2.30\pm0.01$ & $1.10\pm0.02$, $2.70\pm0.01$ \\
$\rm \theta = 60^{\circ}$ & $0.28\pm0.06$, $1.20\pm0.06$ & $0.16\pm0.05$, $0.90\pm0.05$ & $-0.07\pm0.03$, $0.59\pm0.03$ & $0.78\pm0.02$, $2.30\pm0.01$ & $1.10\pm0.02$, $2.70\pm0.01$ \\
$\rm \theta = 75^{\circ}$ & $0.28\pm0.06$, $1.50\pm0.06$ & $0.17\pm0.05$, $1.20\pm0.05$ & $-0.05\pm0.03$, $0.88\pm0.03$ & $0.80\pm0.02$, $2.30\pm0.01$ & $1.10\pm0.02$, $2.70\pm0.01$ \\
\hline
\end{tabular}%
}
\label{tab:eq_mdot_th}
\raggedright
\tablecomments{Tabulated numbers are coefficients in $\log({\rm BC}) = (m\pm \Delta m)\,\log(\lambda_{\rm Edd}) + (c\pm\Delta c)$ for different \emph{inclination} values (averaged over spin). Values are shown for wavelength--inclination combinations with significant correlation ($p<0.05$).}
\end{table*}

\begin{table*}[th!]
    \caption{Comparison of bolometric correction factors}
    \centering
    \resizebox{\textwidth}{!}{%
    \begin{tabular}{c|ccc|cccccc}
    \hline \hline
        Band & \multicolumn{3}{c|}{This paper\tablenotemark{a}}   
        & \citeauthor{Elvis1994} & \citeauthor{Elvis1994} 
        & \citeauthor{Richards2006} & \citeauthor{Richards2006}
        & \citeauthor{R12} & \citeauthor{Nemmen2010} \\
        (rest-frame) & spin = $-1$ & spin = 0 & spin = 1 & (1994) & recalculated by & (2006) & recalculated by & (2012) & (2010) \\
        &   &   &   &   & \citeauthor{R12} &   & \citeauthor{R12} &   &   \\
        \hline
5100~\AA
& $3.4^{6.6}_{2.5}$ & $4.1^{8.3}_{2.5}$ & $11.9^{24.9}_{6.4}$
& 12.5\tablenotemark{b} & 7.7\tablenotemark{c}
& 10.3\tablenotemark{d} & 5.5\tablenotemark{e}
& 8.1 & 7.6 \\

3000~\AA 
& $2.6^{3.6}_{2.0}$ & $2.5^{4.2}_{2.2}$ & $6.3^{12.5}_{3.6}$
& 6.2\tablenotemark{b} & 3.8\tablenotemark{c}
& 5.6\tablenotemark{d} & 3.1\tablenotemark{e}
& 5.2 & 5.9 \\

1450~\AA
& $2.2^{3.1}_{1.8}$ & $2.3^{3.0}_{1.8}$ & $3.0^{5.4}_{2.2}$ 
& 5.1\tablenotemark{b} & 3.2\tablenotemark{c}
& \no & 2.3\tablenotemark{e}
& 4.2 & 3.0 \\

2~keV
& $58.9^{135.7}_{20.5}$ & $58.7^{127.1}_{21.3}$ & $57.5^{88.3}_{28.6}$
& \no & \no
& \no & \no
& 38.0\tablenotemark{f} & \no \\

10~keV
& $71.7^{249.2}_{16.7}$ & $72.4^{238.5}_{17.7}$ & $81.2^{201.5}_{27.1}$
& \no & \no
& \no & \no
& 38.0\tablenotemark{f} & \no \\
\hline
    \end{tabular}}
    \label{tab:wave_bc}
\tablenotetext{a}{\raggedright Values reported are the median and 25th--75th percentile ranges. We report values for an inclination angle of $30^{\circ}$, consistent with the Type~1 AGN assumptions adopted in the studies with which we compare our results. The median values for other inclination angles are reported in Figure~\ref{fig:bc_hist}.} 

\tablenotetext{b}{\raggedright  \citet{Elvis1994} did not present bolometric corrections exactly at these wavelengths (instead presented them at  the $V$, $B$, and 2500~\AA\ bands, see Section~\ref{sec:lit}); the reported values are calculated by \citet{R12} by integrating the radio-to-X-ray mean SED of radio-quiet AGN from \citet{Elvis1994}.} 
\tablenotetext{c}{\raggedright Values calculated by \citet{R12} by integrating the mean SED of the radio-quiet quasars from \citet{Elvis1994} over 1\,\micron--8\,keV, corrected for isotropy.} 

\tablenotetext{d}{\raggedright Values reported by \citet{Richards2006}, calculated by integrating over 100\,\micron--10\,keV. \citeauthor{Richards2006} did not include a bolometric correction at 1450\,\AA.} 
\tablenotetext{e}{\raggedright  Values calculated by \citet{R12} by integrating the mean SEDs of \citet{Richards2006} over 1\,\micron--8\,keV, corrected for inclination angle.}

\tablenotetext{f}{ \citet{R12} report the integrated 2--10 keV bolometric correction for a mix of radio-loud and radio-quiet sample.} \end{table*}


\subsection{Comparison With the Literature at Visible Wavelengths} \label{sec:lit}

Table \ref{tab:wave_bc} compares our bolometric corrections with those reported in the literature. Because all the studies we reference provide bolometric corrections for quasars, i.e., Type~1 AGN,  Table~\ref{tab:wave_bc} reports estimates for an inclination angle of 30$^{\circ}$ to ensure a consistent comparison. 
The comparison studies contain a mix of radio-loud and radio-quiet sources. While the launch of powerful radio jets is generally associated with high spin values, some studies indicate that radio-quiet AGN may also have large spins (e.g., \citealt{lb2006}). Therefore, rather than labeling different spin values as corresponding to radio-loud or radio-quiet populations,  Table~\ref{tab:wave_bc} reports bolometric corrections for three representative spin values (spin\,$=-1, 0, 1$).

\citet{Elvis1994} analyzed a $z<1$ heterogeneous sample of 47 bright quasars (including 18 radio-loud) observed with the \textit{Einstein Observatory}, each with sufficient counts to constrain the soft-X-ray slope. 
The sources were also required to be detected by \textit{IUE} and to have ample multi-wavelength (radio-to-X-ray) photometry available. 
The sample is biased toward blue (unobscured) quasars with strong soft X-ray emission. 
\citet{Elvis1994} reported bolometric corrections at the $V$, $B$, and 2500~\AA\ bands and found $L_{\rm bol}/L_{2500}=5.6$, $L_{\rm bol}/L_{B}=10.7$, and $L_{\rm bol}/L_{V}=13.2$. 
\citet{R12} recalculated the bolometric corrections for the mean SED of the radio-quiet subset of the \citet{Elvis1994} sample. 
In addition, \citet{R12} restricted the integration to 1~$\mu$m--8~keV and applied isotropy corrections:
because accretion disks emit more radiation perpendicular to the disk than in the disk plane, assuming isotropy in a sample viewed primarily pole-on overestimates $L_{\rm bol}$. Adopting an average quasar viewing angle of $\sim$30$^{\circ}$ \citep{Barthel1989}, \citet{R12} recommended scaling $L_{\rm bol} \approx 0.75\,L_{\rm iso}$ and reported bolometric corrections of 12.45 at 5100~\AA, 6.19 at 3000~\AA, and 5.12 at 1450~\AA\  (Table~\ref{tab:wave_bc}).
Our bolometric correction estimations broadly agree with the \citet{Elvis1994} values. However, \citet{Elvis1994} considered a wider integration range (radio through X-rays) than ours.

 \citet{Richards2006} estimated the bolometric correction for a sample of SDSS quasars for which multi-wavelength radio-to-X-ray data were available. Their sample includes both blue and red (intrinsically reddened) Type~1 quasars, and while it contains both radio-loud and radio-quiet sources, it is dominated by radio-quiet quasars. The bolometric corrections were calculated by integrating the SEDs over the range 100\,\micron--10\,keV. \citet{R12} recalculated the bolometric corrections of \citeauthor{Richards2006} by restricting the integration to 1~$\mu$m--8~keV and applying isotropy corrections (given in Table~\ref{tab:wave_bc}). Our median values are broadly consistent with those of \citet{Richards2006}. The differences may arise from the integration ranges—100\,\micron--10\,keV in \citet{Richards2006} versus 1\,\micron--10\,keV in our work, which excludes reprocessed IR emission—as well as from our explicit treatment of inclination effects in the integration. The difference is less pronounced when compared with the \citeauthor{Richards2006} recalculated values.

\citet{R12} studied 63 bright quasars (both radio-loud and radio-quiet) at $z < 1.4$ and calculated the bolometric luminosity by integrating over 1~\micron--8~keV. As noted above, they corrected the bolometric luminosity by assuming a single inclination angle of $\sim$30$^{\circ}$ and adjusted the integrated luminosity accordingly. \citet{R12} found that bolometric corrections at visible–UV wavelengths were similar for the two populations (within the 95\% confidence intervals), but significant differences appeared in the X-ray regime.  As noted above, we avoid classifying AGN as radio-loud or radio-quiet based on spin, but  Table~\ref{tab:wave_bc} shows substantial overlap in the bolometric correction values at each wavelength, including the X-ray regime, for the three tabulated spin values. This is despite \citet{R12} having bridged gaps in SED wavelength coverage by interpolating between the FUV and X-rays, whereas our SED models self-consistently span the full wavelength range.

Our results overall are consistent with the direct estimates of \citet{R12} and their recalculated bolometric corrections for \citet{Richards2006} and \citet{Elvis1994}. However, while they corrected for inclination effects by assuming a single angle, our approach explicitly integrates over the full range of $0^{\circ}$--$90^{\circ}$. 

A factor to consider in studies that estimate bolometric luminosities based on radio-loud samples \citep[e.g.,][]{Elvis1994,Richards2006,R12} is that, when integrating the observed SED up to X-ray energies, the emission may include high-energy contributions from radio structures. In contrast, our SEDs arise purely from the corona and the accretion disk with no contribution from radio structures. \citet{Azadi2020} showed that, in some powerful 3CR radio-loud sources, the contribution from radio structures to the X-ray band can reach up to $\sim$50\% of the total observed flux. Therefore, even if the integration range does not explicitly include the radio regime, X-ray emission from radio structures in radio-loud quasars can still bias the results and lead to an overestimate of the bolometric luminosity.

Unlike \citet{Elvis1994}, \citet{Richards2006}, and \citet{R12}, who determined empirical bolometric corrections from observed SEDs, \citet{Nemmen2010} used the \citet{Hubeny2001} accretion disk model to derive theoretical bolometric corrections. Their parameter space included non-rotating (spin = 0) and maximally rotating (spin = 1) SMBHs with masses varying from $1.25\times10^8$
to $3.2\times10^{10}$~$\rm M_{\odot}$, Eddington ratio within the range of 0.01 to 0.7, and inclination angles varying from face-on to edge-on. While overall our results are consistent within the uncertainties,  differences can be attributed to the difference in parameter space (Table~\ref{tab:qsosed}) and that the \citet{Nemmen2010} model extends only from 3\,\um to 414\,eV and thus does not account for hard-X-ray emission from the corona.


\subsection{Comparison with the Literature at X-ray Wavelengths} \label{sec:lit2}

The bolometric corrections at X-ray energies (last two rows of Table~\ref{tab:wave_bc}) span an enormous range. This is primarily due to the strong and steep dependence of the X-ray bolometric correction on Eddington ratio, as illustrated in Figure~\ref{fig:bc_mdot_wave_spin}. By contrast, the dependence on spin (Table~\ref{tab:wave_bc}) is relatively weak. \citet{R12} estimated X-ray (2--10~keV) bolometric corrections for the radio-loud and the radio-quiet samples (Table~\ref{tab:wave_bc}) and found systematically higher X-ray bolometric corrections for radio-quiet quasars compared to radio-loud ones. This suggests radio loudness is not controlled simply by spin.
Although these \citet{R12} estimates are consistent with ours within the uncertainties, the differences can be attributed to several factors. One is methodological: \citet{R12} bridged the gap between the FUV and X-ray regimes using linear interpolations based on either a power law or the models of \citet{FM1987} and \citet{K1997}. These interpolations may yield different luminosities in this wavelength range compared to the model of \citet{Kubota2018}, thereby producing different bolometric corrections. Another factor is the X-ray contribution from radio jets, which \citet{R12} did not account for.

\citet{Hopkins2007} presented the bolometric luminosity function for quasars at $z<6$ from the MIR to the hard X-ray regime, along with a luminosity-dependent bolometric correction. For quasars with 2\,keV luminosities of $10^{44}$\,erg\,s$^{-1}$, they report a bolometric correction of $\sim$20, whereas we find values ranging from $\sim$10 to a few hundreds, depending on Eddington ratio. The \citet{Hopkins2007} simplified approach to estimating the bolometric luminosity using the visible-to-X-ray spectral index ($\alpha_{\rm ox}$) and the $L_{\nu}(2500\,\text{\AA})$ relation, does not account (as our method does) for intrinsic differences in accretion disk SEDs driven by SMBH, Eddington ratio, and spin (see also \citealt{V2007}).

Using a sample of AGN at $z < 0.7$ that combined Far-Ultraviolet Spectroscopic Explorer (FUSE) UV data with X-ray photometry from the Advanced Satellite for Cosmology and Astrophysics (ASCA), {XMM–Newton}, and {Chandra}, \citet{V2007} found no strong evidence for a luminosity dependence in the bolometric correction but did identify a clear dependence on Eddington ratio, consistent with our results. Specifically, for $\log(\lambda_{\rm Edd}) < -1.0$, they reported X-ray (2--10~keV) bolometric corrections of 15--25, increasing to 40--70 for $\log(\lambda_{\rm Edd}) > -1.0$, with some individual cases exceeding 100. While our values are generally higher, there is substantial overlap between the ranges reported in the two studies, and the studies agree that there is a strong dependence of the X-ray bolometric correction on Eddington ratio.


\subsection{Advantages and Limitations of Determining the Bolometric Correction from accretion disk Modeling}
\label{sec:lim}

Obtaining bolometric luminosities from the QSOSED accretion disk model 
has a number of advantages. One is that our model self-consistently predicts the entire visible to X-ray SED\null.  In contrast, direct observations often have a large gap between the far-UV and X-ray.  Most studies \citep[e.g.,][]{Elvis1994,Richards2006,R12} bridge this gap by linearly interpolating the UV and X-ray observations in $\log\nu$--$\log(\nu L_{\nu})$ space, but this fails to reconstruct correctly the SED peak, where the accretion disk emits most energy. Another advantage of our method is that it accounts purely for the emission from the accretion disk and corona, providing a direct estimate of the accretion luminosity. This distinction is particularly important for radio-loud sources: while visible--UV bolometric corrections still trace the accretion disk reliably, the X-ray band can include significant contributions from jet-related emission. In such cases, using X-ray bolometric corrections without correcting for the jet component risks overestimating the true accretion luminosity.

Powerful AGN are generally assumed to host optically thick, geometrically thin accretion disks \citep[e.g.,][]{SS1976, NT1973} that radiate strongly, but anisotropically, in the visible--UV.
When calculating the radiative power of an AGN, this anisotropy must be taken into account. Some studies correct the observed SEDs by adopting a single average inclination angle for the entire sample \citep[e.g.,][]{R12}; however, this is an oversimplification, and our approach explicitly accounts for inclination by integrating over the full range  ($0^{\circ}$--$90^{\circ}$) of inclination angles.

Relativistic effects such as beaming, aberration, and light bending can produce significant departures from simple Newtonian assumptions, leading to more complex observed spectra. These departures become especially noticeable at large inclination angles \citep[e.g.,][]{Hubeny2001, Nemmen2010}. \citeauthor{Nemmen2010} found that, when relativistic effects are included, the bolometric luminosity along a single line of sight can range from 67\% to 200\% of the integrated isotropic luminosity, for inclination angles $\theta \lesssim 15^{\circ}$ and $\theta \gtrsim 75^{\circ}$ respectively. The QSOSED model does not include relativistic effects such as strong reflection or relativistic smearing to explain the ``soft X-ray excess'' in AGN spectra. Instead, it assumes that the accretion disk truncates at the region where the hard X-ray emission originates \citep[e.g.,][]{Yaqoob2016, Porquet2018} and incorporates a warm Comptonization region to account for the soft X-ray excess. This scenario has been increasingly favored over relativistic disk-reflection models by recent studies \citep[e.g.,][]{Porquet2018}. This work  presents bolometric corrections for inclination angles ranging from 15$^{\circ}$--75$^{\circ}$. However, caution is required when using values for inclinations greater than 45$^{\circ}$. At these orientations, relativistic effects can strongly influence the emission and introduce anisotropies \citep[e.g.,][]{Nemmen2010}. In addition, obscuration from the circumnuclear ``torus'' and the host galaxy becomes significant and must be accounted for when transitioning from an observed SED to an intrinsic one.

As noted in Section~\ref{sec:intro}, there is a disagreement in the literature over the wavelength range of observations that determines AGN  bolometric luminosity. Some studies suggest that integrating over the visible-to-X-ray range provides a reliable estimate  \citep[e.g.,][]{Nemmen2010, R12}. \cite{R12} argued that under the assumption of isotropy, no reprocessed emission should be included in the bolometric luminosity calculation. MIR photons from the torus are nearly all reprocessed visible--UV photons from the accretion disk, which are already accounted for. By contrast, \cite{Richards2006} included the MIR--X-ray emission, arguing that the integrated MIR emission may be a better indicator of bolometric luminosity, given that MIR emission is more isotropic than accretion disk emission. 

A final caveat is that our accretion disk templates are restricted to the parameter space defined in Table~\ref{tab:qsosed}. Consequently, the bolometric correction estimates presented here may not be reliable for AGN with masses, Eddington ratios, or inclination angles that fall outside the ranges considered.

\section{A Guidance for Observers} \label{sec:recipe}

\subsection{Optimal Wavelengths for Estimating Bolometric Luminosity}\label{sec:bestbc}

A key question is which wavelength provides the most robust estimate of the bolometric luminosity. The optimal choice is the band that shows the least sensitivity to SMBH mass, Eddington ratio, spin, and inclination. Our results indicate that the UV band at 1450\,\AA\ is the most reliable, as it lies close to the peak of the accretion disk SED and shows only weak dependence on these parameters. The X-ray band is also a strong indicator, as it depends primarily on the Eddington ratio and is largely independent of the other parameters. Nevertheless, this dependence leads to a wide  range of bolometric corrections in the X-rays, and this must be taken into account when estimating bolometric luminosities. Notably, the independence of the X-ray bolometric correction from most fundamental AGN properties implies that an observer who measures $L_{\rm X}$ can directly apply our bolometric corrections to infer the Eddington ratio using the relations shown in Figures~\ref{fig:bc_mdot_wave_spin} and~\ref{fig:bc_mdot_wave_th}.

\subsection{A Recipe for Observers}

Our SEDs are given in the rest frame and are therefore not restricted by redshift, and the resulting corrections can be applied uniformly to AGN samples across cosmic time. Moreover, because our SEDs are intrinsic, they can be used across the visible-to-X-ray range to estimate fundamental parameters such as SMBH mass and Eddington ratio. Nevertheless, depending on the AGN type, several additional factors must be considered when performing SED fitting or estimating bolometric luminosity.  

\textbf{Case 1: Single-band observations in the visible--X-ray range without prior information:}  
the observed flux at all frequencies from 1~\micron\ to X-rays is affected by dust and gas obscuration along the line of sight. For Type~1 sources, this effect is relatively modest. In that case, Figure~\ref{fig:bc_hist} can be used to infer a bolometric correction. For example, a flux measurement around 3000\,\AA\ compared with Figure~\ref{fig:bc_hist}, assuming $\rm spin=0$ and an inclination of $\sim$30$^{\circ}$, yields a bolometric correction of 2.5. However, when the full range of spin ($-1 \leq\rm spin \leq 1$) and inclination ($\theta \leq 45^{\circ}$) is considered, the bolometric correction can vary between 1.8 and 15.2, with the exact value also depending on SMBH mass and Eddington ratio. A similar situation holds at other wavelengths, including the X-rays. Thus, rather than adopting a single value, observers without prior knowledge of SMBH mass or Eddington ratio should consider the full range of corrections to account for these uncertainties.  

\textbf{Case 2: Multi-band observations in the visible--X-ray range without prior information:}  
Figure~\ref{fig:bc_hist} can also be used to infer bolometric corrections when two or more bands are available. In such cases, we recommend using the bolometric correction at the band closest to the optimal wavelength (see Section \ref{sec:bestbc}).  

\textbf{Case 3: Prior information on SMBH mass or Eddington ratio available:}  
When prior constraints on SMBH mass or Eddington ratio are available, Figures~\ref{fig:new_bc_mass} and~\ref{fig:new_bc_mdot} (also tabulated in Tables~\ref{tab:app3} and \ref{tab:app4}.) can be used to obtain more precise bolometric corrections. Furthermore, multiwavelength observations allow the source to be placed on our SED templates, narrowing the allowed parameter space and yielding more constrained estimates of the bolometric correction.  

\textbf{Caveats.}  
The observed photometry of Type~1 AGN is affected by the dusty torus as well as dust in the host galaxy. This effect can be corrected in several ways: (i) by fitting torus models to the MIR photometry, (ii) by estimating the dust content from the neutral hydrogen column density ($N_{\rm H}$) obtained through X-ray spectral fitting and assuming a gas-to-dust ratio, or (iii) by using the 9.7~$\mu$m silicate feature to estimate the line-of-sight dust optical depth when MIR spectra are available. However, if the torus is clumpy, the silicate feature absorption depth may not be a reliable tracer of the total obscuration even in Type~1 AGN \citep{Siebenmorgen2015}. In Type~2 AGN, or in sources observed at large inclination angles, obscuration from both the torus and the host galaxy is more severe. This increases the uncertainty in SED modeling and in the corresponding estimates of bolometric luminosity.

\section{Summary} \label{sec:summary}

To better understand how AGN impact their host galaxies and surrounding environments (e.g., galaxy clusters), it is essential to determine their total radiative power. This quantity can be estimated from the accretion-disk SED because the accretion disk is the primary source of radiation in AGN\null. This study presents intrinsic accretion-disk SEDs based on the \cite{Kubota2018} QSOSED models along with the corresponding bolometric corrections derived from them. 
The bolometric correction estimates are based on integrating the accretion-disk SEDs from 1\,\micron\ to 10\,keV, taking the inclination-dependence of the emitted flux into account. There is no need to include the MIR emission because photons reprocessed by dust are already accounted for in the visible--UV portion of the SED\null. Additionally, no gap repair is required because our model self-consistently covers the full wavelength range of the accretion-disk and corona emission  (Equation~\ref{eq:bc}). However, the intrinsic SEDs presented here do not account for obscuration by dust in either the torus or the host galaxy. In particular, for Type 2 sources, an external obscuration correction method is required

Our main findings are summarized below.

\begin{itemize}

\item The accretion disk SED---and consequently the bolometric corrections derived from it---depends strongly on SMBH mass and Eddington ratio but only weakly on spin and inclination. Increasing the SMBH mass produces a cooler disk whose SED peak shifts toward lower frequencies, making the effect of mass most pronounced in the visible--NUV bands. By contrast, higher Eddington ratios and/or larger spins yield hotter disks with SED peaks shifted to higher frequencies. At higher inclinations, less visible--UV radiation is observed because the disk subtends a smaller solid angle, and the near side of the disk can obscure regions farther away. In contrast, the X-ray emission is largely isotropic and essentially independent of inclination.

\item The bolometric correction shows distinct parameter dependencies across different wavelengths. In the visible--NUV bands (5100\,\AA\ and 3000\,\AA) show a strong dependence on SMBH mass. At X-ray energies, the bolometric correction depends primarily on the Eddington ratio but is largely independent of spin, SMBH mass, and inclination. Although SMBH mass still affects the visible–UV range, its impact is weaker than the strong Eddington ratio dependence observed in the X-rays (Figures~\ref{fig:bc_mass_wave_spin}–\ref{fig:bc_mdot_wave_th}).

\item Dependence on either  SMBH mass or Eddington ratio is weaker near the peak of the accretion disk SED (FUV\, $\sim$1450\,\AA; Figures~\ref{fig:bc_mass_wave_spin} and~\ref{fig:bc_mdot_wave_spin}), implying that bolometric corrections at FUV wavelengths are less sensitive to the underlying accretion disk parameters. Consequently, the FUV band at 1450\,\AA\ provides the most robust estimate of the bolometric correction.

\item For observers with single- or multi-band data, a practical recipe for applying our bolometric corrections is presented in Section~\ref{sec:recipe}. For observers with only a single-band observation and no prior knowledge of their AGN properties, we recommend using Figure~\ref{fig:bc_hist}. The full range of bolometric correction factors should be taken into account, particularly in the X-rays, where the spread is larger due to the strong dependence on the Eddington ratio. A further caveat is the impact of dust in the torus and host galaxy, which becomes more significant at higher inclination angles. Observers with prior information on SMBH mass and Eddington ratio can make use of the more detailed dependencies shown in Figure~\ref{fig:new_bc_mass} and Figure~\ref{fig:new_bc_mdot}, along with the tabulated values in Tables~\ref{tab:app3} and \ref{tab:app4}.

\end{itemize}

The SED templates presented here are applicable to AGN at any redshift.  An observer with photometry in the rest-frame range from 1\,\um to 10\,keV after correcting for obscuration along the line of sight (due to dust and gas in the line-of-sight through the host galaxy and/or torus) can apply these SEDs or their corresponding bolometric corrections to estimate the intrinsic radiative power of their sources.

\section*{Acknowledgements}

Support for this work was provided by NASA grants: \#80NSSC18K1609,  \#80NSSC19K1311,
 \#80NSSC20K0043  (MAz), and \#80NSSC21K0058 (JK), and by NASA Contract NAS8-03060   \textit{Chandra} X-ray Center (CXC), which is operated by the Smithsonian Astrophysical Observatory   
(BJW, JK). BJW acknowledges the support of the Royal Society and
the Wolfson Foundation at the University of Bristol.

The scientific results in this article are based to a significant degree on observations made by the \textit{Chandra}~X-ray Observatory (CXO).
This research has made use of data obtained from the \textit{Chandra} Data
Archive. This research is based on observations made by {\it Herschel}, which
is an ESA space observatory with science instruments provided by European-led Principal Investigator consortia and with important
participation from NASA. This work is based in part on observations made with the Spitzer Space
Telescope, which was operated by the Jet Propulsion Laboratory, California Institute of Technology under a contract with NASA.

We acknowledge the use of Ned Wright's calculator
\citep{2006PASP..118.1711W} and NASA/IPAC Extragalactic Database (NED), operated by the Jet Propulsion Laboratory, California Institute
of Technology, under contract with the National Aeronautics and Space
Administration.

The authors would like to thank Chris Done, Mark Birkinshaw, and Diana Worrall for helpful comments that improved the quality of the paper.

\bibliography{references.bib}

\setlength{\LTpre}{0pt}
\setlength{\LTpost}{2pt}
\setlength{\LTleft}{0pt}
\setlength{\LTright}{0pt}
\setlength{\aboverulesep}{0pt}
\setlength{\belowrulesep}{0pt}

\appendix
\restartappendixnumbering   
\section{SED Template Tables}
\centering

{\tiny
\setlength{\tabcolsep}{5.00pt}
\renewcommand{\arraystretch}{0.70}
\setlength{\aboverulesep}{0pt}
\setlength{\belowrulesep}{0pt}
\setlength{\LTpre}{0pt}
\setlength{\LTpost}{2pt}
\setlength{\LTleft}{0pt}
\setlength{\LTright}{0pt}
\begin{longtable}{@{\hskip 2pt}r@{\hskip 2pt}c@{\hskip 2pt}c@{\hskip 2pt}c@{\hskip 2pt}c@{\hskip 2pt}c@{\hskip 2pt}c@{\hskip 2pt}c@{\hskip 2pt}c@{\hskip 2pt}c@{\hskip 2pt}c@{\hskip 2pt}c@{\hskip 2pt}c@{\hskip 2pt}c@{\hskip 2pt}c@{\hskip 2pt}c@{\hskip 2pt}c@{\hskip 2pt}c@{\hskip 2pt}c@{\hskip 2pt}c@{\hskip 2pt}c@{\hskip 2pt}c@{\hskip 2pt}c@{\hskip 2pt}c@{}}
\caption{ The numerical values for the SED templates in Figure \ref{fig:new_ad_mass}} \label{tab:app1} \\
\toprule
\bfseries Num & \multicolumn{4}{c}{\bfseries Params} & \multicolumn{19}{c}{\bfseries $\log \nu$ [rest-frame]} \\
 \\ 
\bfseries & $\log \lambda_{\mathrm{Edd}}$ & $\log (M_{\rm BH}/M_\odot)$ & \bfseries a & $\theta\,[^{\circ}]$ & \tiny 14.00 & \tiny 14.25 & \tiny 14.50 & \tiny 14.75 & \tiny 15.00 & \tiny 15.25 & \tiny 15.50 & \tiny 15.75 & \tiny 16.00 & \tiny 16.25 & \tiny 16.50 & \tiny 16.75 & \tiny 17.00 & \tiny 17.25 & \tiny 17.50 & \tiny 17.75 & \tiny 18.00 & \tiny 18.25 & \tiny 18.50 \\
\midrule
\endfirsthead
\toprule
\bfseries Num & \multicolumn{4}{c}{\bfseries Params} & \multicolumn{19}{c}{\bfseries $\log \nu$ [rest-frame]} \\
 \\ 
\bfseries & $\log \lambda_{\mathrm{Edd}}$ & $\log (M_{\rm BH}/M_\odot)$ & \bfseries a & $\theta\,[^{\circ}]$ & \tiny 14.00 & \tiny 14.25 & \tiny 14.50 & \tiny 14.75 & \tiny 15.00 & \tiny 15.25 & \tiny 15.50 & \tiny 15.75 & \tiny 16.00 & \tiny 16.25 & \tiny 16.50 & \tiny 16.75 & \tiny 17.00 & \tiny 17.25 & \tiny 17.50 & \tiny 17.75 & \tiny 18.00 & \tiny 18.25 & \tiny 18.50 \\
\midrule
\endhead
\midrule
\multicolumn{24}{r}{\footnotesize\itshape continued on next page} \\
\endfoot
\bottomrule
\endlastfoot
1 & -1.5 & 7--8 & -1.0 & 15 & 42.8 & 42.8 & 43.0 & 43.5 & 43.6 & 43.6 & 43.4 & 43.2 & 43.2 & 43.1 & 43.0 & 42.9 & 42.9 & 42.8 & 42.8 & 42.9 & 43.0 & 43.0 & 43.1 \\
2 & -1.5 & 8--9 & -1.0 & 15 & 44.1 & 44.1 & 44.3 & 44.6 & 44.5 & 44.3 & 44.2 & 44.1 & 44.0 & 44.0 & 43.9 & 43.9 & 43.8 & 43.8 & 43.8 & 43.9 & 44.0 & 44.0 & 44.1 \\
3 & -1.5 & 9--10 & -1.0 & 15 & 45.2 & 45.2 & 45.4 & 45.5 & 45.3 & 45.2 & 45.1 & 45.0 & 44.9 & 44.9 & 44.8 & 44.8 & 44.8 & 44.8 & 44.8 & 44.9 & 45.0 & 45.0 & 45.1 \\
4 & -1.5 & 7--8 & -1.0 & 30 & 42.8 & 42.8 & 43.0 & 43.4 & 43.6 & 43.5 & 43.3 & 43.2 & 43.1 & 43.0 & 43.0 & 42.9 & 42.9 & 42.8 & 42.8 & 42.9 & 43.0 & 43.0 & 43.1 \\
5 & -1.5 & 8--9 & -1.0 & 30 & 44.1 & 44.1 & 44.2 & 44.5 & 44.5 & 44.3 & 44.2 & 44.1 & 44.0 & 43.9 & 43.9 & 43.8 & 43.8 & 43.8 & 43.8 & 43.9 & 44.0 & 44.0 & 44.1 \\
6 & -1.5 & 9--10 & -1.0 & 30 & 45.2 & 45.2 & 45.3 & 45.4 & 45.3 & 45.2 & 45.1 & 45.0 & 44.9 & 44.8 & 44.8 & 44.8 & 44.8 & 44.8 & 44.8 & 44.9 & 45.0 & 45.0 & 45.1 \\
7 & -1.5 & 7--8 & -1.0 & 45 & 42.7 & 42.7 & 42.9 & 43.3 & 43.5 & 43.4 & 43.2 & 43.1 & 43.0 & 43.0 & 42.9 & 42.9 & 42.8 & 42.8 & 42.8 & 42.9 & 43.0 & 43.0 & 43.1 \\
8 & -1.5 & 8--9 & -1.0 & 45 & 44.0 & 44.0 & 44.1 & 44.4 & 44.4 & 44.2 & 44.1 & 44.0 & 43.9 & 43.9 & 43.8 & 43.8 & 43.8 & 43.8 & 43.8 & 43.9 & 44.0 & 44.0 & 44.1 \\
\end{longtable}
}

\vspace{2cm}

{\tiny
\setlength{\tabcolsep}{5.00pt}
\renewcommand{\arraystretch}{0.70}
\setlength{\aboverulesep}{0pt}
\setlength{\belowrulesep}{0pt}
\setlength{\LTpre}{0pt}
\setlength{\LTpost}{2pt}
\setlength{\LTleft}{0pt}
\setlength{\LTright}{0pt}
\begin{longtable}{@{\hskip 2pt}r@{\hskip 2pt}c@{\hskip 2pt}c@{\hskip 2pt}c@{\hskip 2pt}c@{\hskip 2pt}c@{\hskip 2pt}c@{\hskip 2pt}c@{\hskip 2pt}c@{\hskip 2pt}c@{\hskip 2pt}c@{\hskip 2pt}c@{\hskip 2pt}c@{\hskip 2pt}c@{\hskip 2pt}c@{\hskip 2pt}c@{\hskip 2pt}c@{\hskip 2pt}c@{\hskip 2pt}c@{\hskip 2pt}c@{\hskip 2pt}c@{\hskip 2pt}c@{\hskip 2pt}c@{\hskip 2pt}c@{}}
\caption{ The numerical values for the SED templates in Figure \ref{fig:new_ad_mdot}} \label{tab:app2} \\

\toprule
\bfseries Num & \multicolumn{4}{c}{\bfseries Params} & \multicolumn{19}{c}{\bfseries $\log \nu$ [rest-frame]} \\
 \\ 
\bfseries & $\log \lambda_{\mathrm{Edd}}$ & $\log (M_{\rm BH}/M_\odot)$ & \bfseries a & $\theta\,[^{\circ}]$ & \tiny 14.00 & \tiny 14.25 & \tiny 14.50 & \tiny 14.75 & \tiny 15.00 & \tiny 15.25 & \tiny 15.50 & \tiny 15.75 & \tiny 16.00 & \tiny 16.25 & \tiny 16.50 & \tiny 16.75 & \tiny 17.00 & \tiny 17.25 & \tiny 17.50 & \tiny 17.75 & \tiny 18.00 & \tiny 18.25 & \tiny 18.50 \\
\midrule
\endfirsthead
\toprule
\bfseries Num & \multicolumn{4}{c}{\bfseries Params} & \multicolumn{19}{c}{\bfseries $\log \nu$ [rest-frame]} \\
 \\ 
\bfseries & $\log \lambda_{\mathrm{Edd}}$ & $\log (M_{\rm BH}/M_\odot)$ & \bfseries a & $\theta\,[^{\circ}]$ & \tiny 14.00 & \tiny 14.25 & \tiny 14.50 & \tiny 14.75 & \tiny 15.00 & \tiny 15.25 & \tiny 15.50 & \tiny 15.75 & \tiny 16.00 & \tiny 16.25 & \tiny 16.50 & \tiny 16.75 & \tiny 17.00 & \tiny 17.25 & \tiny 17.50 & \tiny 17.75 & \tiny 18.00 & \tiny 18.25 & \tiny 18.50 \\
\midrule
\multicolumn{24}{r}{\footnotesize\itshape continued on next page} \\
\endfoot
\bottomrule
\endlastfoot
1 & -1.5 -- -1.0 & 7 & -1.0 & 15 & $42.3$ & $42.3$ & $42.6$ & $43.1$ & $43.4$ & $43.6$ & $43.6$ & $43.4$ & $43.2$ & $43.1$ & $43.0$ & $42.9$ & $42.8$ & $42.7$ & $42.6$ & $42.5$ & $42.6$ & $42.6$ & $42.6$ \\
2 & -1.0 -- -0.5 & 7 & -1.0 & 15 & $42.7$ & $42.7$ & $43.0$ & $43.4$ & $43.8$ & $44.0$ & $44.2$ & $44.2$ & $44.0$ & $43.8$ & $43.7$ & $43.5$ & $43.4$ & $43.2$ & $42.9$ & $42.7$ & $42.7$ & $42.6$ & $42.6$ \\
3 & -0.5 -- 0.0 & 7 & -1.0 & 15 & $43.1$ & $43.1$ & $43.3$ & $43.8$ & $44.1$ & $44.4$ & $44.6$ & $44.8$ & $44.8$ & $44.5$ & $44.2$ & $44.0$ & $43.9$ & $43.6$ & $43.2$ & $42.8$ & $42.7$ & $42.7$ & $42.6$ \\
4 & -1.5 -- -1.0 & 7 & -1.0 & 30 & $42.3$ & $42.3$ & $42.5$ & $43.0$ & $43.3$ & $43.5$ & $43.5$ & $43.4$ & $43.2$ & $43.1$ & $43.0$ & $42.9$ & $42.8$ & $42.7$ & $42.6$ & $42.5$ & $42.6$ & $42.6$ & $42.6$ \\
5 & -1.0 -- -0.5 & 7 & -1.0 & 30 & $42.7$ & $42.7$ & $42.9$ & $43.4$ & $43.7$ & $44.0$ & $44.2$ & $44.2$ & $44.0$ & $43.7$ & $43.6$ & $43.5$ & $43.3$ & $43.1$ & $42.8$ & $42.7$ & $42.7$ & $42.6$ & $42.6$ \\
6 & -0.5 -- 0.0 & 7 & -1.0 & 30 & $43.1$ & $43.1$ & $43.3$ & $43.7$ & $44.1$ & $44.4$ & $44.6$ & $44.7$ & $44.7$ & $44.4$ & $44.2$ & $44.0$ & $43.8$ & $43.6$ & $43.2$ & $42.8$ & $42.7$ & $42.7$ & $42.6$ \\
7 & -1.5 -- -1.0 & 7 & -1.0 & 45 & $42.2$ & $42.2$ & $42.4$ & $42.9$ & $43.2$ & $43.4$ & $43.4$ & $43.3$ & $43.1$ & $43.0$ & $42.9$ & $42.8$ & $42.7$ & $42.6$ & $42.6$ & $42.5$ & $42.6$ & $42.6$ & $42.6$ \\
8 & -1.0 -- -0.5 & 7 & -1.0 & 45 & $42.6$ & $42.6$ & $42.8$ & $43.3$ & $43.6$ & $43.9$ & $44.1$ & $44.1$ & $43.9$ & $43.7$ & $43.5$ & $43.4$ & $43.3$ & $43.1$ & $42.8$ & $42.7$ & $42.7$ & $42.6$ & $42.6$ \\
\end{longtable}
}
\vspace{2cm}
\section{Bolometric Correction Tables}
\vspace{2cm}

{\tiny
\setlength{\tabcolsep}{5.00pt}
\renewcommand{\arraystretch}{0.70}
\setlength{\aboverulesep}{0pt}
\setlength{\belowrulesep}{0pt}
\setlength{\LTpre}{0pt}
\setlength{\LTpost}{2pt}
\setlength{\LTleft}{0pt}
\setlength{\LTright}{0pt}
\begin{longtable}{@{\hskip 2pt}r@{\hskip 2pt}c@{\hskip 2pt}c@{\hskip 2pt}c@{\hskip 2pt}c@{\hskip 2pt}c@{\hskip 2pt}c@{\hskip 2pt}c@{\hskip 2pt}c@{\hskip 2pt}c@{\hskip 2pt}c@{\hskip 2pt}c@{\hskip 2pt}c@{\hskip 2pt}c@{\hskip 2pt}c@{\hskip 2pt}c@{\hskip 2pt}c@{\hskip 2pt}c@{\hskip 2pt}c@{\hskip 2pt}c@{\hskip 2pt}c@{\hskip 2pt}c@{\hskip 2pt}c@{\hskip 2pt}c@{}}
\caption{The numerical values for the bolometric corrections in Figure \ref{fig:new_bc_mass}} \label{tab:app3} \\
\toprule
\bfseries Num & \multicolumn{4}{c}{\bfseries Params} & \multicolumn{19}{c}{\bfseries $\log \nu$ [rest-frame]} \\
 \\ 
\bfseries & $\log \lambda_{\mathrm{Edd}}$ & $\log (M_{\rm BH}/M_\odot)$ & \bfseries a & $\theta\,[^{\circ}]$ & \tiny 14.00 & \tiny 14.25 & \tiny 14.50 & \tiny 14.75 & \tiny 15.00 & \tiny 15.25 & \tiny 15.50 & \tiny 15.75 & \tiny 16.00 & \tiny 16.25 & \tiny 16.50 & \tiny 16.75 & \tiny 17.00 & \tiny 17.25 & \tiny 17.50 & \tiny 17.75 & \tiny 18.00 & \tiny 18.25 & \tiny 18.50 \\
\midrule
\endfirsthead
\toprule
\bfseries Num & \multicolumn{4}{c}{\bfseries Params} & \multicolumn{19}{c}{\bfseries $\log \nu$ [rest-frame]} \\
 \\ 
\bfseries & $\log \lambda_{\mathrm{Edd}}$ & $\log (M_{\rm BH}/M_\odot)$ & \bfseries a & $\theta\,[^{\circ}]$ & \tiny 14.00 & \tiny 14.25 & \tiny 14.50 & \tiny 14.75 & \tiny 15.00 & \tiny 15.25 & \tiny 15.50 & \tiny 15.75 & \tiny 16.00 & \tiny 16.25 & \tiny 16.50 & \tiny 16.75 & \tiny 17.00 & \tiny 17.25 & \tiny 17.50 & \tiny 17.75 & \tiny 18.00 & \tiny 18.25 & \tiny 18.50 \\
\midrule
\endhead
\midrule
\multicolumn{24}{r}{\footnotesize\itshape continued on next page} \\
\endfoot
\bottomrule
\endlastfoot
1 & -1.5 & 7--8 & -1.0 & 15 & 14.5 & 14.5 & 8.99 & 3.28 & 2.09 & 2.49 & 3.98 & 5.16 & 6.40 & 7.84 & 9.32 & 10.7 & 11.9 & 12.9 & 13.1 & 11.6 & 9.81 & 8.26 & 7.55 \\
2 & -1.5 & 8--9 & -1.0 & 15 & 6.76 & 6.76 & 4.50 & 2.24 & 2.37 & 3.79 & 4.91 & 6.17 & 7.60 & 9.08 & 10.5 & 11.6 & 12.3 & 12.7 & 12.4 & 10.8 & 9.13 & 7.69 & 7.03 \\
3 & -1.5 & 9--10 & -1.0 & 15 & 4.27 & 4.27 & 3.24 & 2.46 & 3.40 & 4.41 & 5.54 & 6.92 & 8.34 & 9.68 & 10.8 & 11.5 & 11.7 & 11.6 & 10.9 & 9.40 & 7.94 & 6.68 & 6.10 \\
4 & -1.5 & 7--8 & -1.0 & 30 & 16.2 & 16.2 & 10.0 & 3.65 & 2.33 & 2.77 & 4.42 & 5.68 & 7.00 & 8.50 & 9.99 & 11.4 & 12.4 & 13.2 & 13.2 & 11.6 & 9.81 & 8.26 & 7.55 \\
5 & -1.5 & 8--9 & -1.0 & 30 & 7.54 & 7.54 & 5.02 & 2.50 & 2.64 & 4.20 & 5.42 & 6.77 & 8.27 & 9.78 & 11.1 & 12.1 & 12.7 & 12.9 & 12.4 & 10.8 & 9.13 & 7.69 & 7.03 \\
6 & -1.5 & 9--10 & -1.0 & 30 & 4.77 & 4.77 & 3.61 & 2.74 & 3.78 & 4.87 & 6.08 & 7.54 & 9.01 & 10.3 & 11.4 & 11.9 & 12.0 & 11.7 & 10.9 & 9.40 & 7.94 & 6.68 & 6.10 \\
7 & -1.5 & 7--8 & -1.0 & 45 & 19.8 & 19.8 & 12.3 & 4.47 & 2.85 & 3.38 & 5.35 & 6.78 & 8.22 & 9.81 & 11.3 & 12.5 & 13.3 & 13.6 & 13.3 & 11.6 & 9.81 & 8.26 & 7.55 \\
8 & -1.5 & 8--9 & -1.0 & 45 & 9.24 & 9.24 & 6.15 & 3.06 & 3.22 & 5.10 & 6.49 & 7.99 & 9.60 & 11.1 & 12.4 & 13.1 & 13.4 & 13.3 & 12.5 & 10.8 & 9.13 & 7.69 & 7.03 \\
\end{longtable}
}

\vspace{2cm}

{\tiny
\setlength{\tabcolsep}{5.00pt}
\renewcommand{\arraystretch}{0.70}
\setlength{\aboverulesep}{0pt}
\setlength{\belowrulesep}{0pt}
\setlength{\LTpre}{0pt}
\setlength{\LTpost}{2pt}
\setlength{\LTleft}{0pt}
\setlength{\LTright}{0pt}
\begin{longtable}{@{\hskip 2pt}r@{\hskip 2pt}c@{\hskip 2pt}c@{\hskip 2pt}c@{\hskip 2pt}c@{\hskip 2pt}c@{\hskip 2pt}c@{\hskip 2pt}c@{\hskip 2pt}c@{\hskip 2pt}c@{\hskip 2pt}c@{\hskip 2pt}c@{\hskip 2pt}c@{\hskip 2pt}c@{\hskip 2pt}c@{\hskip 2pt}c@{\hskip 2pt}c@{\hskip 2pt}c@{\hskip 2pt}c@{\hskip 2pt}c@{\hskip 2pt}c@{\hskip 2pt}c@{\hskip 2pt}c@{\hskip 2pt}c@{}}
\caption{The numerical values for the bolometric corrections in Figure \ref{fig:new_bc_mdot}} \label{tab:app4} \\
\toprule
\bfseries Num & \multicolumn{4}{c}{\bfseries Params} & \multicolumn{19}{c}{\bfseries $\log \nu$ [rest-frame]} \\
 \\ 
\bfseries & $\log \lambda_{\mathrm{Edd}}$ & $\log (M_{\rm BH}/M_\odot)$ & \bfseries a & $\theta\,[^{\circ}]$ & \tiny 14.00 & \tiny 14.25 & \tiny 14.50 & \tiny 14.75 & \tiny 15.00 & \tiny 15.25 & \tiny 15.50 & \tiny 15.75 & \tiny 16.00 & \tiny 16.25 & \tiny 16.50 & \tiny 16.75 & \tiny 17.00 & \tiny 17.25 & \tiny 17.50 & \tiny 17.75 & \tiny 18.00 & \tiny 18.25 & \tiny 18.50 \\
\midrule
\endfirsthead
\toprule
\bfseries Num & \multicolumn{4}{c}{\bfseries Params} & \multicolumn{19}{c}{\bfseries $\log \nu$ [rest-frame]} \\
 \\ 
\bfseries & $\log \lambda_{\mathrm{Edd}}$ & $\log (M_{\rm BH}/M_\odot)$ & \bfseries a & $\theta\,[^{\circ}]$ & \tiny 14.00 & \tiny 14.25 & \tiny 14.50 & \tiny 14.75 & \tiny 15.00 & \tiny 15.25 & \tiny 15.50 & \tiny 15.75 & \tiny 16.00 & \tiny 16.25 & \tiny 16.50 & \tiny 16.75 & \tiny 17.00 & \tiny 17.25 & \tiny 17.50 & \tiny 17.75 & \tiny 18.00 & \tiny 18.25 & \tiny 18.50 \\
\midrule
\endhead
\midrule
\multicolumn{24}{r}{\footnotesize\itshape continued on next page} \\
\endfoot
\bottomrule
\endlastfoot
1 & -1.50 -- -1.00 & 7 & -1.0 & 15 & $31.8$ & $31.8$ & $19.3$ & $6.28$ & $3.01$ & $2.01$ & $1.89$ & $2.94$ & $4.30$ & $5.35$ & $6.75$ & $8.56$ & $10.9$ & $14.5$ & $19.2$ & $20.0$ & $18.6$ & $17.3$ & $16.6$ \\
2 & -1.00 -- -0.50 & 7 & -1.0 & 15 & $46.9$ & $46.9$ & $28.8$ & $9.64$ & $4.47$ & $2.48$ & $1.63$ & $1.53$ & $2.43$ & $4.26$ & $5.64$ & $7.64$ & $10.9$ & $17.9$ & $35.5$ & $52.1$ & $56.0$ & $59.5$ & $61.6$ \\
3 & -0.50 -- 0.00 & 7 & -1.0 & 15 & $62.9$ & $62.9$ & $39.1$ & $13.6$ & $6.33$ & $3.32$ & $1.92$ & $1.37$ & $1.49$ & $2.91$ & $5.36$ & $7.54$ & $11.2$ & $20.4$ & $55.2$ & $125$ & $155$ & $187$ & $207$ \\
4 & -1.50 -- -1.00 & 7 & -1.0 & 30 & $35.4$ & $35.4$ & $21.5$ & $7.00$ & $3.35$ & $2.24$ & $2.11$ & $3.27$ & $4.74$ & $5.86$ & $7.33$ & $9.21$ & $11.5$ & $15.1$ & $19.4$ & $20.0$ & $18.6$ & $17.3$ & $16.6$ \\
5 & -1.00 -- -0.50 & 7 & -1.0 & 30 & $52.3$ & $52.3$ & $32.1$ & $10.8$ & $4.99$ & $2.76$ & $1.82$ & $1.70$ & $2.71$ & $4.71$ & $6.19$ & $8.35$ & $11.8$ & $19.1$ & $36.6$ & $52.2$ & $56.0$ & $59.5$ & $61.6$ \\
6 & -0.50 -- 0.00 & 7 & -1.0 & 30 & $70.1$ & $70.1$ & $43.6$ & $15.2$ & $7.06$ & $3.70$ & $2.15$ & $1.52$ & $1.66$ & $3.24$ & $5.92$ & $8.29$ & $12.3$ & $22.2$ & $58.1$ & $126$ & $155$ & $187$ & $207$ \\
7 & -1.50 -- -1.00 & 7 & -1.0 & 45 & $43.4$ & $43.4$ & $26.3$ & $8.58$ & $4.11$ & $2.74$ & $2.58$ & $3.98$ & $5.67$ & $6.89$ & $8.49$ & $10.5$ & $12.8$ & $16.1$ & $19.8$ & $20.0$ & $18.6$ & $17.3$ & $16.6$ \\
8 & -1.00 -- -0.50 & 7 & -1.0 & 45 & $64.1$ & $64.1$ & $39.3$ & $13.2$ & $6.11$ & $3.38$ & $2.23$ & $2.08$ & $3.30$ & $5.65$ & $7.33$ & $9.79$ & $13.6$ & $21.4$ & $38.4$ & $52.3$ & $56.0$ & $59.5$ & $61.6$ \\
\end{longtable}
}

\end{document}